\documentclass[journal,twoside]{IEEEtran}
\usepackage{amssymb}
\usepackage{amsthm}
\usepackage{multirow}
\usepackage{inputenc}
\usepackage{enumerate} 
\usepackage{textcomp}
\usepackage{listings}
\usepackage{array}
\usepackage[switch,mathlines]{lineno}
\usepackage{soul}
\usepackage{xfrac}
\usepackage{graphicx}
\usepackage{cite}
\usepackage[cmex10]{amsmath}
\usepackage{mathrsfs}
\usepackage{algorithm}
\usepackage{xcolor}
\usepackage{colortbl}
\usepackage{makecell}
\usepackage{cancel}
\usepackage{algorithm,algpseudocode}
\algnewcommand{\Inputs}[1]{%
  \State \textbf{Inputs:}
  \Statex \hspace*{\algorithmicindent}\parbox[t]{.8\linewidth}{\raggedright #1}
}
\algnewcommand{\Initialize}[1]{%
  \State \textbf{Initialize:}
  \Statex \hspace*{\algorithmicindent}\parbox[t]{.8\linewidth}{\raggedright #1}
}
\makeatletter
\def\footnoterule{\kern-3\p@
  \hrule \@width 3.3in \kern 2.6\p@} 
\makeatother

\usepackage{algpseudocode}

\usepackage{stackengine}

\usepackage{algpseudocode}
\usepackage{CJK}
\usepackage[cmex10]{amsmath}
\usepackage{bm}
\usepackage{color}
\usepackage{amsmath,amsthm,amssymb,amsfonts}
\usepackage{flushend}
\usepackage{float}
\newtheorem{theorem}{Theorem}

\newtheorem{remark}{Remark}

\usepackage{arydshln} 
\usepackage{hyperref}
\hypersetup{
     colorlinks   = true,
     linkcolor    = blue,
     citecolor    = red,
     urlcolor     = blue
}
\makeatletter
\newcommand*{\transpose}{%
  {\mathpalette\@transpose{}}%
}
\newcommand*{\@transpose}[2]{%
  \raisebox{\depth}{$\m@th#1\intercal$}%
}
\makeatother

\usepackage{soul}
\usepackage{tikz}
\usepackage{booktabs}
\usepackage{tabularx}

\usepackage[caption=false,font=footnotesize]{subfig}
\ifCLASSINFOpdf
\else
\fi

\begin{document}
\renewcommand{\ttdefault}{cmtt}
\bstctlcite{IEEEexample:BSTcontrol}

\title{ Matrix Low-dimensional Qubit Casting Based Quantum Electromagnetic Transient Network Simulation Program}



\author{
{Qi Lou,~\IEEEmembership{Student Member},
Yijun~Xu,~\IEEEmembership{Senior Member}, 
Wei Gu,~\IEEEmembership{Senior Member},
}
\thanks{Q. Lou, Y. Xu and W. Gu are with the school of Electrical Engineering, Southeast University, Nanjing, Jiangsu 210096, China (e-mail: \{\texttt{louqi, yijunxu, wgu\}@seu.edu.cn}).

This work was supported by the National Natural Science Foundation of China under Grant 52325703.

\emph{(Corresponding author: Wei Gu.)}}
}

\maketitle
\begin{abstract}
In modern power systems, the integration of converter-interfaced generations requires the development of electromagnetic transient network simulation programs (EMTP) that can capture rapid fluctuations. 
However, as the power system scales, the EMTP's computing complexity increases exponentially, leading to a curse of dimensionality that hinders its practical application. 
Facing this challenge, quantum computing offers a promising approach for achieving exponential acceleration. 
To realize this in noisy intermediate-scale quantum computers, the variational quantum linear solution (VQLS) was advocated because of its robustness against depolarizing noise. However, it suffers data inflation issues in its preprocessing phase, and no prior research has applied quantum computing to high-frequency switching EMT networks.
To address these issues, this paper first designs the matrix low-dimension qubit casting (MLQC) method to address the data inflation problem in the preprocessing of the admittance matrix for VQLS in EMT networks. Besides,  we propose a real-only quantum circuit reduction method tailored to the characteristics of the EMT network admittance matrices. Finally, the proposed quantum EMTP algorithm (QEMTP) has been successfully verified for EMT networks containing a large number of high-frequency switching elements. 
\end{abstract}


\IEEEpeerreviewmaketitle
 \vspace{-0.2cm}

\section{Introduction}
\IEEEPARstart{R}{ecently}, the increasing penetration of renewables challenges the secure operation of modern power systems. Its associated fast-switching components in power electronic devices further lead to multi-timescale features in its dynamics behaviors, which cannot be well captured by traditional electromechanical transient simulations programs\cite{watson2003power}. Consequently, EMTP is gaining increasing popularity because of its ability to track the dynamics of the converter-dominated power system accurately.
However, achieving fast solutions for EMTP remains a significant challenge since the binary classical computer exhibits extremely high computing complexity as the dimension of the nonlinear equation in EMTP increases\cite{electrical_power_systems, TQE2}. This is especially true for large-scale power systems with various power electronic devices.

Luckily, with the rapid advancement of quantum computers that have natural parallelism and entanglement properties, there is hope that this bottleneck can be overcome\cite{TQE1,TQE3}. Therefore, some researchers explored the feasibility of quantum EMTP(QEMTP). For example, Zhou \emph{et al.} proposed a Harrow-Hassidim-Lloyd(HHL)-based QEMTP, marking the first proof-of-concept attempt to apply quantum solutions to EMTP \cite{QEMTP}. However, the HHL algorithm requires a substantial quantum circuit depth that increases with the size of the problem scale \cite{HHL_ori}. Using it, even a simple RLC series circuit requires $7$ qubits and $102$ circuit layers, which is an unmanageable depth for today’s quantum computers.
Besides, this excessive depth makes it highly sensitive to noise that is impractical for today's noisy intermediate-scale quantum computers\cite{TQE4,Quantum_computation_in_power_systems,NISQ2018quantum,cerezo2021variational}. Therefore, it is no surprise that this HHL-based QEMTP is only tested in a noise-free quantum simulator \cite{QEMTP}.

To enable the QEMTP algorithm to operate in noisy intermediate-scale quantum devices, Zhou \emph{et al.} \cite{zhou2022noisy} further proposed the VQLS-based QEMTP. As a classical-quantum hybrid algorithm, VQLS transforms the linear system of equations into an unconstrained optimization problem, where the quantum computer evaluates the cost function while the classical computer performs the optimization. Theoretically, VQLS only requires variational quantum circuits with shallow depth and logarithmic width, making it more resistant to noise and more efficient use of quantum resources \cite{vqanoise}. The approach was tested successfully on IBM's real quantum hardware, achieving promising results. However, as a trade-off. The VQLS achieves less computing acceleration compared to the HHL algorithm \cite{Quantum_computation_in_power_systems}. Furthermore, in practice, the VQLS-based QEMTP method still faces the following challenges:
\begin{enumerate}
    \item VQLS-based QEMTP suffers from unacceptable admittance matrix preprocessing time since the decomposition of the coefficient matrix involves computational complexity that increases exponentially with matrix size, resulting in impractically long preprocessing times.
    \item The VQLS algorithm requires numerous quantum circuits to run in parallel to obtain a single cost function result, while the circuit number is polynomially related to the count of unitary matrix decompositions. As the problem size increases, even a simple circuit requires a large number of parallel circuits that overburden the quantum computing resources.
    \item In VQLS, the admittance matrix needs to be reprocessed if there is a topology change, making it difficult to apply in high-speed power electronic devices with rapid switching topologies. That is why in \cite{zhou2022noisy}, the case study was limited to IEEE standard networks that lack power electronic devices, whose transient analysis rarely demands a time scale below the microsecond level.
\end{enumerate}

Facing the above challenges, this paper presents a matrix low-dimensional qubit casting (MLQC)-based method for QEMTP. 
It yields the following contributions:
\begin{itemize}
    \item An MLQC-based preprocessing method for the admittance matrix is proposed. Using physical relationships among multiple qubits to achieve low-dimensional decomposition of the matrix, this approach effectively reduces the preprocessing time required for the VQLS algorithm.
    \item A real-only quantum circuit construction method is proposed. By integrating the physical characteristics of the admittance matrix in EMT networks, this method effectively reduces the number of quantum circuits required to compute the cost function, significantly saving parallel quantum computing resources.
    \item QEMTP is investigated, for the first time, in high-frequency switching EMT networks using the proposed fixed admittance switch model (FASM) that successfully considers power electronic devices.
\end{itemize}
The simulation results reveal excellent performance of the proposed method. 

The remainder of the paper is organized as follows. Section.\ref{Framework} introduces the QEMTP framework, Section.\ref{VQLS-Based QEMTP solver} summarizes the VQLS-Based QEMTP, Section.\ref{section_MLQC} presents the proposed MLQC-based QEMTP, Section.\ref{case_study} provides case studies, followed by the conclusion in Section.\ref{conclu}.

\section{QEMTP algorithm Framework}\label{Framework}
First, the preliminaries for our work is provided, including the EMTP and QEMTP formulation.

\subsection{Formulations of the EMTP}
\begin{figure}[!htbp]
  \begin{center}
    \includegraphics[scale=0.4]{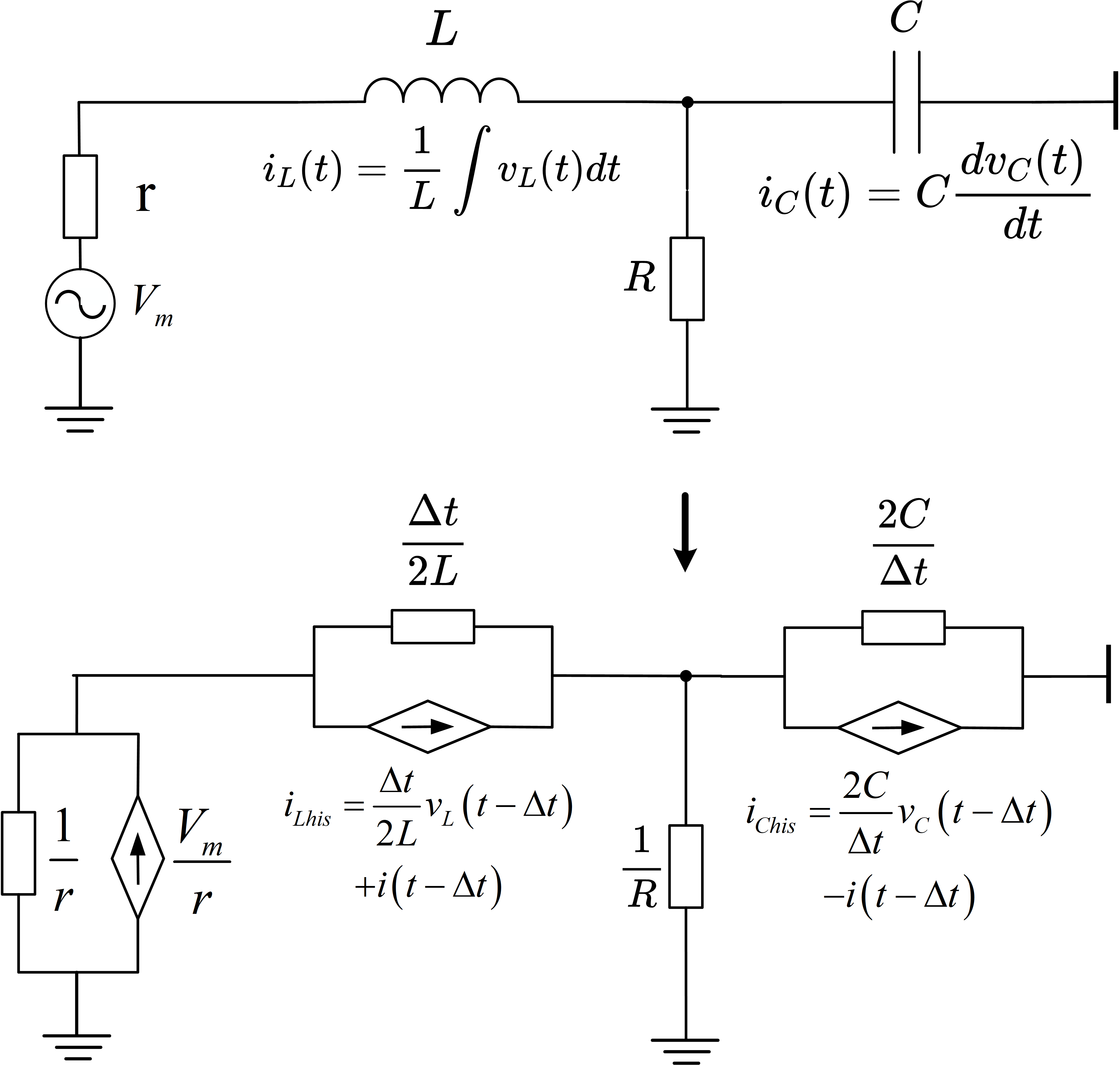}
    \setlength{\abovecaptionskip}{-0pt}
    \vspace{-0cm}
    \caption{The EMTP model for RLC components. By discretizing the voltage and current equations of each electrical component in the network using the trapezoidal method, the final EMTP equivalent network can be constructed for the basic RLC components. }
    \label{fig:1}
  \end{center}
\end{figure}

Traditionally, the EMTP replaces circuit elements with their Norton equivalents and formulates the associated branch current and node voltage equations \cite{EMTPBOOK}. By discretizing the electrical components using numerical methods, e.g. the trapezoidal integration method, the Runge-Kutta method, etc., the elements can be transformed into parallel configurations of equivalent resistors and current sources. An illustrative example of basic RLC components is shown in Fig \ref{fig:1}. 

Then, after applying the Norton equivalent to each electrical component in the network, the node equations of the network can be formulated as
\begin{equation}\label{EMTP}
     \mathbf{G}\vec{v}(t) = \vec{i}_{inj}(t)+\vec{I}_{history}\equiv \vec{i}(t).
\end{equation}
Here, $\mathbf{G}$ denotes the network admittance matrix, $\Vec{v}(t)$ denotes the vector of nodal voltages in the network, $\vec{i}(t)$ denotes the vector of nodal injected current vector, $\vec{i}_{inj}(t)$ denotes the externally injected current in the current time step, and $\Vec{I}_{history}$ denotes the vector of historical nodal injected current. Once the network node equation is established, the voltage and current of the network at each time step can be obtained through iterative loops. 

\subsection{Formulations of the QEMTP}
Once the classical EMTP algorithm is constructed, it needs to be mapped to the Hilbert space since quantum computing uses quantum states as a fundamental unit. In quantum computers, since each qubit is a two-dimensional vector and multiple qubits are related through the Kronecker product, the dimension of the corresponding space for $n$ qubits is $2^{n}$ as \cite{nielsen2010quantum}

\begin{equation}\label{mulqubits}
     \left | \Psi  \right \rangle 
=\left | \psi_{1}  \right \rangle \otimes \dots \otimes\left | \psi_{n}  \right \rangle
=   \sum_{j=1}^{2^{n}} \alpha_{j}\left | j  \right \rangle.
\end{equation}

Here, $\left | \Psi  \right \rangle $ denotes a set of $n$ qubits; $\left | \psi_{j}  \right \rangle$ denotes a single qubit in the set $\left | \Psi  \right \rangle $; $\left | j  \right \rangle$ denotes the j-th basis, that is, from $\left | 0\dots 0  \right \rangle$ to $\left | 1\dots 1  \right \rangle$, which is the binary representation from 0 to $2^{n}-1$;  $\alpha_{i}$ denotes the probability amplitude, satisfying $\sum_{j}\left |\alpha_{j}  \right | ^{2}=1$, $\otimes$ denotes kronecker product. 

Therefore, for $\mathbf{G}$ with a dimension less than $2^{n}$, an extension is needed to match the dimensionality of the quantum space. To avoid the singularity of the extended matrix, $\mathbf{\tilde{G}}$, an identity matrix is added along the diagonal. Thus, for the $N$-dimensional matrix, $\mathbf{G}$, a quantized equivalent admittance matrix can be constructed as
\begin{equation}\label{quantum_G}
\mathbf{\tilde{G}}=
\left[\begin{array}{cc}
\mathbf{G}^{N \times N} & 0 \\
0 & \mathbf{I}^{\left(2^{\left \lceil  \log_{2}{N} \right \rceil }-N\right) \times\left(2^{\left \lceil  \log_{2}{N} \right \rceil }-N\right)}
\end{array}\right].
\end{equation}

Accordingly, $\vec{v}(t)$ and $\vec{i}(t)$ are extended to $2^{\left \lceil  \log_{2}{N} \right \rceil }$-dimensions as
\begin{equation}\label{quantum_vi}
\hat{v}=\begin{bmatrix}
 \vec{v}(t)\\0\end{bmatrix} , \hat{i} =  \begin{bmatrix}
 \vec{i}(t)\\0\end{bmatrix}.
\end{equation}

For the nodal voltage and current vectors, since the amplitude of the basis is probability amplitude, it is necessary to normalize the vectors to project them onto Hilbert space as
\begin{subequations}\label{vi-norm}
  \begin{align}
    |i\rangle&=\frac{\sum_{j} \hat{i}_{j}|j\rangle}{\| \sum_{j} \hat{i}_{j}|j\rangle \|_{2}}\\
    |v\rangle&=\frac{\sum_{j} \hat{v}_{j}|j\rangle}{\| \sum_{j} \hat{v}_{j}|j\rangle \|_{2}}.
  \end{align}
\end{subequations}
Here, $i_{j}$ denotes the $j$-th element of the injected augmented nodal current vector $\hat{i}$, $v_{j}$ denotes the $j$-th element of the augmented nodal voltage vector $\hat{v}$. Ultimately, the quantum EMTP nodal equation can be formulated as
\begin{equation}\label{QEMTP_frame}
\mathbf{\tilde{G}}|v\rangle = |i\rangle.
\end{equation}
Using this model, we will next present the way to solve in a quantum computing manner. 

\section{The Traditional VQLS-Based QEMTP solver}\label{VQLS-Based QEMTP solver}
\begin{figure}[!htbp]
  \begin{center}
    \includegraphics[scale=0.66]{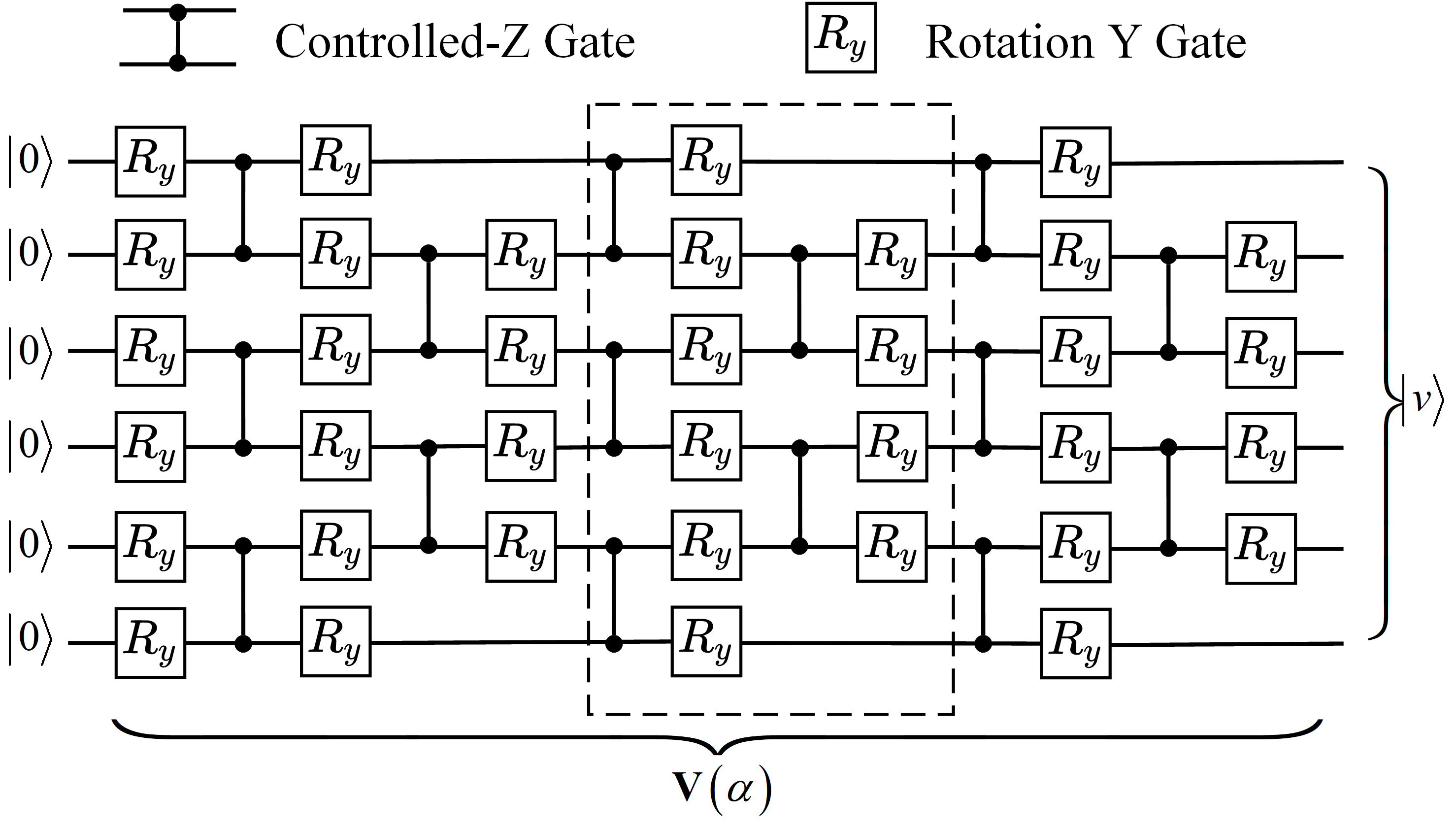}
    \setlength{\abovecaptionskip}{-0pt}
    \vspace{-0cm}
    \caption{Plot for the structure of the hardware-efficient variational quantum circuit. The figure presents the ansatz circuit with qubits number $n=6$. The structure inside the dashed box represents one ansatz layer that includes controlled-Z gates and rotation gates $R_{y}$. By optimizing the rotation angle parameters $\alpha$ in gates $R_{y}$, the circuit can express the desired quantum state. The performances of the circuit can be improved by stacking multiple ansatz layers. The circuit shown in the figure consists of $3$ ansatz layers in series.
}
    \label{fig:ansatz}
  \end{center}
\end{figure}

To solve \eqref{QEMTP_frame}, the VQLS algorithm was advocated for its robustness against noise in real quantum computers. For a linear system of equations as \eqref{QEMTP_frame}, VQLS optimizes the distance between $\mathbf{\tilde{G}}|v\rangle$ and $|i\rangle$ to obtain the final solution $|v\rangle$. By introducing a variational quantum circuit, VQLS transforms $|v\rangle$ into a trainable circuit as \cite{cerezo2021variational}.
\begin{equation}\label{x_ansatz}
|v\rangle = \mathbf{V}(\alpha)|0\rangle.
\end{equation}
Here, $\mathbf{V}(\alpha)$ denotes an ansatz structure, which is a typical quantum circuit structure constructed using parameterized rotation gates and controlled two-qubit gates \cite{wu2021towards}. Then, a hardware-efficient variational quantum circuit can be employed with low circuit depth and high expressive power \cite{kandala2017hardware}. Fig. \ref{fig:ansatz} illustrates its structure\footnote{The controlled-Z (CZ) gate is a two-qubit gate that applies the Pauli-Z gate to the second qubit conditioned on the state of the first qubit. The rotation gate $R_y$ is a single-qubit rotation gate $R_y(\theta) = 
\begin{bmatrix}
\cos\left(\frac{\theta}{2}\right) &  -\sin\left(\frac{\theta}{2}\right) \\
\sin\left(\frac{\theta}{2}\right) &  \cos\left(\frac{\theta}{2}\right)
\end{bmatrix}$
}. By training the parameters, $\alpha$, in the variational circuit, $\mathbf{V}(\alpha)$, the solution, $|i\rangle$, to the linear system can be obtained. More specifically, the implementation of the VQLS-based QEMTP includes four steps, namely admittance matrix projection, cost function construction, optimal parameter training, and error compensation, which will be elaborated next.

\subsection{Admittance Matrix Projection}\label{Matrix Projection}
First, to map the quantized equivalent admittance matrix $\mathbf{\tilde{G}}$ onto a quantum circuit, it is necessary to perform a multi-qubit Pauli gate mapping in Hilbert space. The specific projection method is
\begin{equation}\label{pauli-mapping}
\mathbf{\tilde{G}} = \sum_{i=0}^{4^{n}}c_{i}\mathbf{g}_{i}.
\end{equation}
Here, $\mathbf{g}_{i}\in\otimes_{k=1}^{n}\{\sigma_{I,X,Y,Z}\}$, denotes the $i$-th matrix basis in the set of multi-qubits Pauli gates\footnote{The four  Pauli gates correspond mathematically to the Pauli operators given as
$  \sigma_{I} = \begin{bmatrix}1 & 0\\ 0 & 1\end{bmatrix}\  \sigma_{X} = \begin{bmatrix}0 & 1\\ 1 & 0\end{bmatrix}\ 
  \sigma_{Y} = \begin{bmatrix}0 & -i\\ i & 0\end{bmatrix}\ \sigma_{Z} = \begin{bmatrix}1 & 0\\ 0 & -1\end{bmatrix}.$} corresponding to the high-dimensional Hilbert space constructed by $n$ qubits. $c_{i}$ denotes the projection coefficient of matrix $\mathbf{\tilde{G}}$, obtained by 
\begin{equation}\label{ci_cal}
c_{i}=\frac{1}{2^{n}} T_{r}\left(\mathbf{\tilde{G}} \times \mathbf{g}_{i}\right).
\end{equation}

Then, a complete orthogonal matrix base in a $2^{n}$-dimensional space can be constructed in quantum circuits. Next, we need to encode the current vector, $|i\rangle$. By using a top-down encoding approach \cite{Araujo_2023_encoding}, the amplitude encoding circuit for $|i\rangle$ is derived  as a quantum operator as
\begin{equation}\label{i_encode}
|i\rangle = \mathbf{U}|0\rangle.
\end{equation}
Here, $\mathbf{U}$ is a quantum circuit that represents a unitary matrix. Using it, the information of the node injection current $\vec{i}$ can be encoded in the amplitudes of the quantum state.
\begin{remark}
         Although only 10 qubits are required for a $1024$-dimensional admittance matrix, the number of Pauli-bases exceeds $1$ million. This requires more than $1$ million large-scale matrix calculations for the basis mapping, resulting in a heavy computing burden. For example, this mapping takes over $7$ hours in Python’s NumPy library. Furthermore, in EMT networks with high-frequency power electronic devices, rapid switching causes frequent changes in the admittance matrix. Each new admittance matrix requires recalculating the matrix projection, further deteriorating the computing efficiency. 
    These motivate us to develop the MLQC strategy and an FASM to address this issue, which will be described in Section \ref{section_MLQC}. 
\end{remark}

\subsection{Cost Function Construction}
\begin{figure}[!htbp]
  \begin{center}
    \includegraphics[scale=0.6]{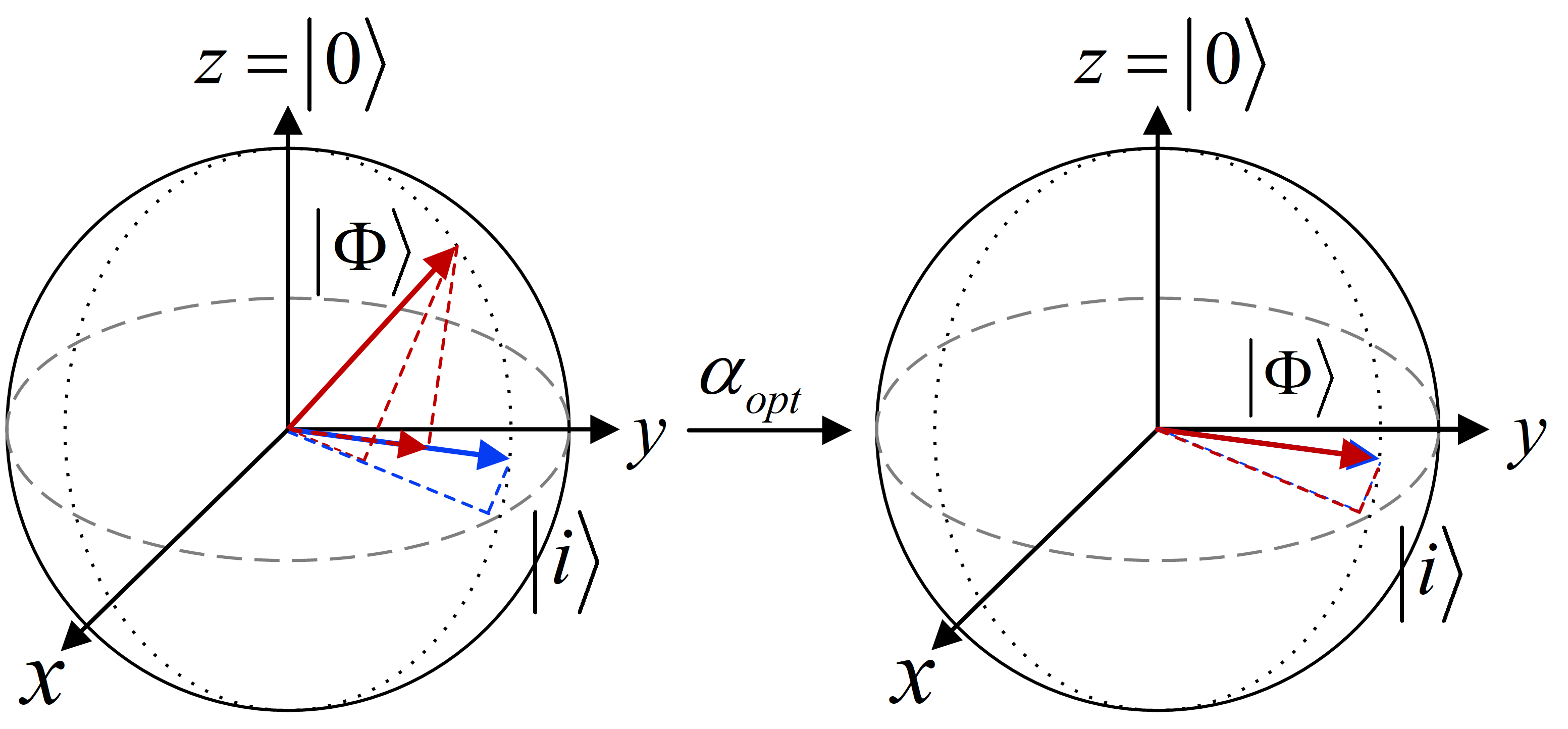}
    \setlength{\abovecaptionskip}{-0pt}
    \vspace{-0cm}
    \caption{The Bloch sphere representation of the projection of $|\Phi\rangle$ onto $|i\rangle$ in the case of a single qubit. The red line represents the initial, unoptimized state $|\Phi\rangle$, while the blue line represents the target state $|i\rangle$. In the beginning, the projection distance between two quantum states is very small. When $|\Phi\rangle$ and $|i\rangle$ are perpendicular, the projection distance is zero. After optimizing the parameter $\alpha$ in the variational circuit, $|\Phi\rangle$ and $|i\rangle$ nearly overlap under the optimal parameters $\alpha_{opt}$, and the projection distance becomes 1. 
}
    \label{fig:2}
  \end{center}
\end{figure}

Next, a cost function is constructed to quantify the similarity between $\mathbf{\tilde{G}}|v\rangle$ and $|i\rangle$. Let $\mathbf{\tilde{G}}|v\rangle$ be $|\phi \rangle$ and its normalized form be $|\Phi\rangle$. When examining the projection distance of $|\Phi\rangle$ on $|i\rangle$, it is observed that the projection is minimized to 0 when $|\Phi\rangle$ and $|i\rangle$ are orthogonal, while maximized to 1 when they are identical. Fig. \ref{fig:2} shows the Bloch sphere visualization where both $|\Phi\rangle$ and $|i\rangle$ are single qubit.

\begin{figure}[!htbp]
  \centering
  \includegraphics[scale=0.66]{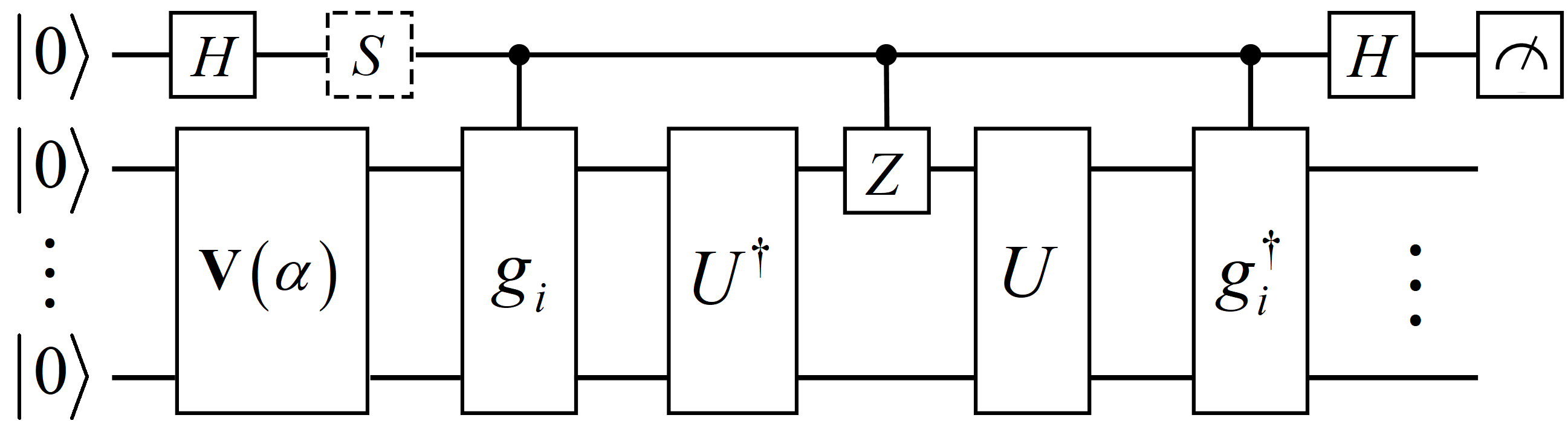}
  
  \setlength{\abovecaptionskip}{0pt} 
  \caption{ 
    Plot for the Hadamard test quantum circuit for calculating $\delta_{ii^{\prime}j}$ under the case of $j=1$. The circuit calculates the expectation value of the Hamiltonian by measuring the probability information of the first qubit, a.k.a. auxiliary qubit. Specifically, the real part of $\delta_{ii^{\prime}j}$ is determined by $P(0)-P(1)$. To compute the imaginary part, an $S$-gate is added to the auxiliary qubit. Then, the same procedure for the real part is conducted. For the calculation of $\beta_{ii^{\prime}}$, it is only necessary to remove the unitary $U$ and $Z_{j}$ gate from the circuit.
  }
  
  \label{fig:cost_delta}
\end{figure}

Now, since the difference between $1$ and the square of the projection distance of $|\Phi\rangle$ on $|i\rangle$ can represent the magnitude of the cost function, 
considering the "barren plateau" problem\cite{cerezo2021cost}\cite{TQE5} which is similar to gradient decent, the trainable local cost function is \cite{bravo2023variational}

\begin{subequations}\label{had_local_cost}
\begin{align}
 C_{L}(\alpha)&=\frac{1}{2} -\frac{1}{2n}\frac{\sum_{j=1}^{n}\sum_{ii^{\prime}j}c_{i}c_{i^{\prime}}^{\ast}\delta_{ii^{\prime}j}}{\sum_{j=1}^{n}\sum_{ii^{\prime} }c_{i}c_{i^{\prime}}^{\ast}\beta_{ii^{\prime}}}\label{cl_delta&beta}
 \\
\delta_{ii^{\prime}j} &= \langle0| \mathbf{V}(\alpha)^{\dagger} \mathbf{g}_{i^{\prime}}^{\dagger} \mathbf{U}\left(Z_{j} \otimes \mathbf{I}_{\bar{j}}\right) \mathbf{U}^{\dagger} \mathbf{g}_{i} \mathbf{V}(\alpha)|0\rangle\label{delta}
\\
\beta_{ii^{\prime}} &= \langle0| \mathbf{V}(\alpha)^{\dagger} \mathbf{g}_{i^{\prime}}^{\dagger} \mathbf{g}_{i} \mathbf{V}(\alpha)|0\rangle\label{beta}
\end{align}
\end{subequations}
Here, $Z_{j}$ represents the application of the Pauli-Z gate to the $j$-th qubit, $\dagger$ denotes the conjugate transpose. By performing parallel computing for $\beta_{ii^{\prime}}$ and 
$\delta_{ii^{\prime}j}$ within the quantum circuit,  the value of the cost function can be ultimately calculated in \eqref{cl_delta&beta}. Here, the Hadamard test circuit can be directly used to calculate the value of $\beta_{ii^{\prime}}$ and $\delta_{ii^{\prime}j}$\cite{aharonov2006polynomial} as shown in Fig. \ref{fig:cost_delta}\footnote{The $S$-gate is a single-qubit phase rotation gate represented as \[\mathbf{S} = \begin{bmatrix} 1 & 0 \\ 0 & i\end{bmatrix}. \]}.

\subsection{Optimal Parameter Training}
Now, the VQLS algorithm optimizes the parameters $\alpha$ to be as close to zero as possible. This optimization can be achieved using gradient descent-based methods as
\begin{equation}\label{grad_des}
\alpha_{k+1} = \alpha_{k}-\eta \nabla C_{L}(\alpha).
\end{equation}
Here, $\alpha_{k}$ represents the parameters in the $k$ th iteration, $\eta$ is the learning rate, and $\nabla C_{L}(\alpha)$ is the cost function gradient with respect to the parameters, $\alpha_{k}$ . For quantum circuits, the gradient is calculated using the parameter shift rule \cite{grad_mitarai2018quantum,grad_schuld2019evaluating}. For a Hadamard test circuit with ansatz shown in Fig. \ref{fig:ansatz}, the gradient of $i$-th parameter $\alpha_{i}$ is given by

\begin{equation}\label{para_shift}
\frac{\partial L(\alpha)}{\partial \alpha_{i}}=\frac{L\left(\alpha_{i}+\frac{\pi}{2}\right)-L\left(\alpha_{i}-\frac{\pi}{2}\right)}{2}.
\end{equation}
Here, $L$ represents the Hadamard test circuit. 
Ultimately, using the optimized parameter $\alpha_{opt}$, the final node voltage $|v\rangle$ can be obtained through \eqref{x_ansatz}.
\subsection{Error Compensation}
Due to limitations in quantum computer coherence time, quantum circuit depth, and noise in quantum computers, the estimated results of the VQLS algorithm contain errors to some extent. Therefore, QEMTP requires an iterative error compensation procedure. After performing \eqref{x_ansatz}, let the quantum state tomography\cite{quantum_state_tomography} of $|v\rangle$ be denoted as $\tilde{v}$. Then, the error, $\Delta i$, at the $k$th iteration can be calculated as
\begin{equation}\label{delta_v}
\Delta i^{(k)}=\hat{i}-\mathbf{\tilde{G}} \tilde{v}^{(k-1)}
\end{equation}
Using $\Delta {i}^{(k)}$, the VQLS algorithm can solve \eqref{compensate_alg} to obtain the correction term.
\begin{equation}\label{compensate_alg}
\mathbf{\tilde{G}} \Delta v^{(k)}=\Delta i^{(k)}
\end{equation}
Here, $\Delta v^{(k)}$ represents the correction voltage obtained in the $k$-th iteration. Based on this correction, the required accuracy is evaluated. The convergence of this algorithm has been proved in \cite{zhou2022noisy}. If the preset accuracy threshold $\epsilon$ is not met, the process is repeated iteratively until the desired error tolerance is achieved. The specific termination condition is given in
\begin{equation}\label{compen_break}
\left\|\mathbf{G} \tilde{v}^{(k)}-\hat{i}\right\|_{2} \leq \epsilon.
\end{equation}
Till now, we have finished the presentation of the VQLS algorithm to solve the QEMTP. 

\section{Our Proposed MLQC-based QEMTP}\label{section_MLQC}
This section proposes an MLQC-based QEMTP algorithm to address the significant time overhead in the admittance matrix projection process mentioned in Section \ref{Matrix Projection}. The proposed method consists of three main procedures: MLQC, Real-only Quantum Circuit Reduction Method, and FASM.
\subsection{The proposed MLQC procedure}
To accelerate the matrix mapping process that VOLS suffers from, we resort to tensor decomposition techniques. This allows us to decompose the matrix into a sum of Kronecker products of several pairs of low-dimensional matrix \cite{van1993approximation}, thus achieving a lossless MLQC.
\subsubsection{Traditional Naive Kronecker Product Decomposition (NKD)}\label{Sec:NKD}
For $\mathbf{\tilde{G}}\in \mathbf{R}^{m\times n}$ with $m=m_{1}m_{2}$ and $n=n_{1}n_{2}$, the naive Kronecker product approximation is formulated as
\begin{equation}\label{Kronecker_app}
\underset{\mathbf{B}, \mathbf{C}}{\arg \min }\left\|\mathbf{\tilde{G}}-\mathbf{B} \otimes \mathbf{C}\right\|_{F}.
\end{equation}
Here, $\mathbf{B}\in \mathbf{R}^{m_{1}\times n_{1}}$ and $\mathbf{C}\in \mathbf{R}^{m_{2}\times n_{2}}$. 

The optimization problem in \eqref{Kronecker_app} can be solved via the Eckhart-Young theorem\cite{EK_theo}. The specific solution process is described in the Appendix.\ref{append:NKD}.

\subsubsection{Generalized Kronecker Product Decomposition (GKD)}
Although NKD can achieve a dimension reduction in the matrix $\mathbf{\tilde{G}}$, it leads to a certain loss of precision. This is surely unsuitable for the EMTP since its admittance matrices typically require extremely high precision. Therefore, we propose investigating the GKD algorithm \cite{hameed2022convolutional}.

Similarly, by modifying the optimization for the  matrix set $\left\{\mathbf{B}_{r}, \mathbf{C}_{r}\right\}_{r=1}^{R}$, the GKD is formulated as

\begin{equation}\label{general_Kronecker_app}
\underset{\left\{\mathbf{B}_{r}, \mathbf{C}_{r}\right\}_{r=1}^{R}}{\arg \min }\left\|\mathbf{\tilde{G}}-\sum_{r=1}^{R}\mathbf{B}_{r} \otimes \mathbf{C}_{r}\right\|_{F}.
\end{equation}
 Here, $\mathbf{\tilde{G}}\in \mathbf{R}^{m_{1}m{2}\times n_{1}n_{2}}$, $\mathbf{B}_{r}\in \mathbf{R}^{m_{1}\times n_{1}}$ and $\mathbf{C}_{r}\in \mathbf{R}^{m_{2}\times n_{2}}$, $R$ is the  parameter that determines the degree of approximation. Then, applying the permute method mentioned in Appendix.\ref{append:NKD}, \eqref{general_Kronecker_app} can be transformed as
 \begin{equation}\label{ge_Kronecker_eq}
\underset{\left\{\mathbf{B}_{r}, \mathbf{C}_{r}\right\}_{r=1}^{R}}{\arg \min }\left\|{\mathbf{\bar{\tilde{G}}}}-\sum_{r=1}^{R}\operatorname{vec}(\mathbf{B}_{r}) \operatorname{vec}(\mathbf{C}_{r})^{T}\right\|_{F}
\end{equation}

Now, using the Eckhart-Young theorem, we have
\begin{equation}\label{gene_ans}
\begin{split}
    \operatorname{vec}(\mathbf{B}_{r})&=\sqrt{\sigma_{r}}\boldsymbol{u}_{r}\\
    \operatorname{vec}(\mathbf{C}_{r})&=\sqrt{\sigma_{r}}\boldsymbol{v}_{r}.
\end{split}
\end{equation}
Here, $\sigma_{r}$ is the $r$-th largest singular value of ${\mathbf{\bar{\tilde{G}}}}$, and $\boldsymbol{u}_{r}$, $\boldsymbol{v}_{r}$ are the corresponding left and right singular vectors.
By increasing the value of $R$, we can improve the accuracy of the approximation. In fact, when $R$ reaches its maximum value $\min \{m_{1} n_{1}, m_{2} n_{2}\}$, $\sum_{r=1}^{R}\operatorname{vec}(\mathbf{B}_{r}) \operatorname{vec}(\mathbf{C}_{r})^{T}$ becomes the SVD form of $\mathbf{\bar{\tilde{G}}}$. Therefore, a \textbf{lossless} Kronecker decomposition can be achieved.

\subsubsection{The MLQC Method}

Through GKD, $\mathbf{\bar{\tilde{G}}}$ is decomposed into multiple sets of Kronecker products of lower-dimensional matrices. This is analogous to the entanglement between two qubits shown in \eqref{mulqubits}, which allows a direct computation of the Pauli basis mapping on the decomposed submatrices. Then, the final mapped results can be obtained with detailed procedures as:

\begin{itemize} 
\item  Using GKD,  $\hat{\mathbf{G}}$ is decomposed into a series of submatrices $\{\mathbf{\Gamma}_{r},\mathbf{Z}_{r}\}_{r=1}^{R}$. We define $\hat{\mathbf{G}}$ as 
\begin{equation}\label{G_decomp}
\hat{\mathbf{G}}=\sum_{r=1}^{R}\mathbf{\Gamma}_{r}\otimes\mathbf{Z}_{r}.
\end{equation}

For an $N$-dimensional matrix which corresponding to $n$ qubits, the dimensions of the submatrices are typically split as $2^{\lceil\frac{n}{2}\rceil}$ and $2^{\lfloor\frac{n}{2}\rfloor}$\footnote{\(\lceil \cdot \rceil\) represents the ceiling function (rounding up), and \(\lfloor \cdot \rfloor\) represents the floor function (rounding down).
}, respectively.

\item Projecting the submatrices $\mathbf{\Gamma}_{r}$ and $\mathbf{Z}_{r}$ onto the pauli operator space, we have
\begin{equation}\label{G_decomp_cast}
\sum_{r=1}^{R}\mathbf{\Gamma}_{r}\otimes\mathbf{Z}_{r} = \sum_{r=1}^{R}(\sum_{i=1}^{2^{\lceil\frac{n}{2}\rceil}}c_{ir}\mathbf{\gamma}_{i}\otimes\sum_{j=1}^{2^{\lfloor\frac{n}{2}\rfloor}}c_{jr}\mathbf{\zeta}_{j}).
\end{equation}
Here, $\mathbf{\gamma}_{i}\in\otimes_{k=1}^{2^{\lceil\frac{n}{2}\rceil}}\{\sigma_{I,X,Y,Z}\}$ and $\mathbf{\zeta}_{j}\in\otimes_{k=1}^{2^{\lfloor\frac{n}{2}\rfloor}}\{\sigma_{I,X,Y,Z}\}$.
\item Expanding \eqref{G_decomp_cast} and combining the projection values with the same basis, we have 
\begin{equation}\label{G_decomp_final}
\hat{\mathbf{G}}= \sum_{i,j}\sum_{r=1}^{R}c_{ir}c_{jr}(\mathbf{\gamma}_{i}\otimes\mathbf{\zeta}_{j}) = \sum_{k=1}^{4^{n}}c_{k}\mathbf{g}_{k}.
\end{equation}
\end{itemize}

Next, we need to proof that the mapping obtained from \eqref{G_decomp_final} under the MLQC method is exactly the same as \eqref{ci_cal}. In the Pauli operator space defined within the Hilbert space, the following theorem holds:
\begin{theorem}[Uniqueness of the matrix mapping]\label{mapping1}
For $\forall \mathbf{\tilde{G}} \in  \mathbf{R}^{2^{n}\times 2^{n}}$, its mapping within the Pauli space formed by $\otimes_{k=1}^{n}\{\sigma_{I,X,Y,Z}\}$ is unique.
\end{theorem}

\begin{figure}[!htbp]
  \begin{center}
    \includegraphics[scale=1.1]{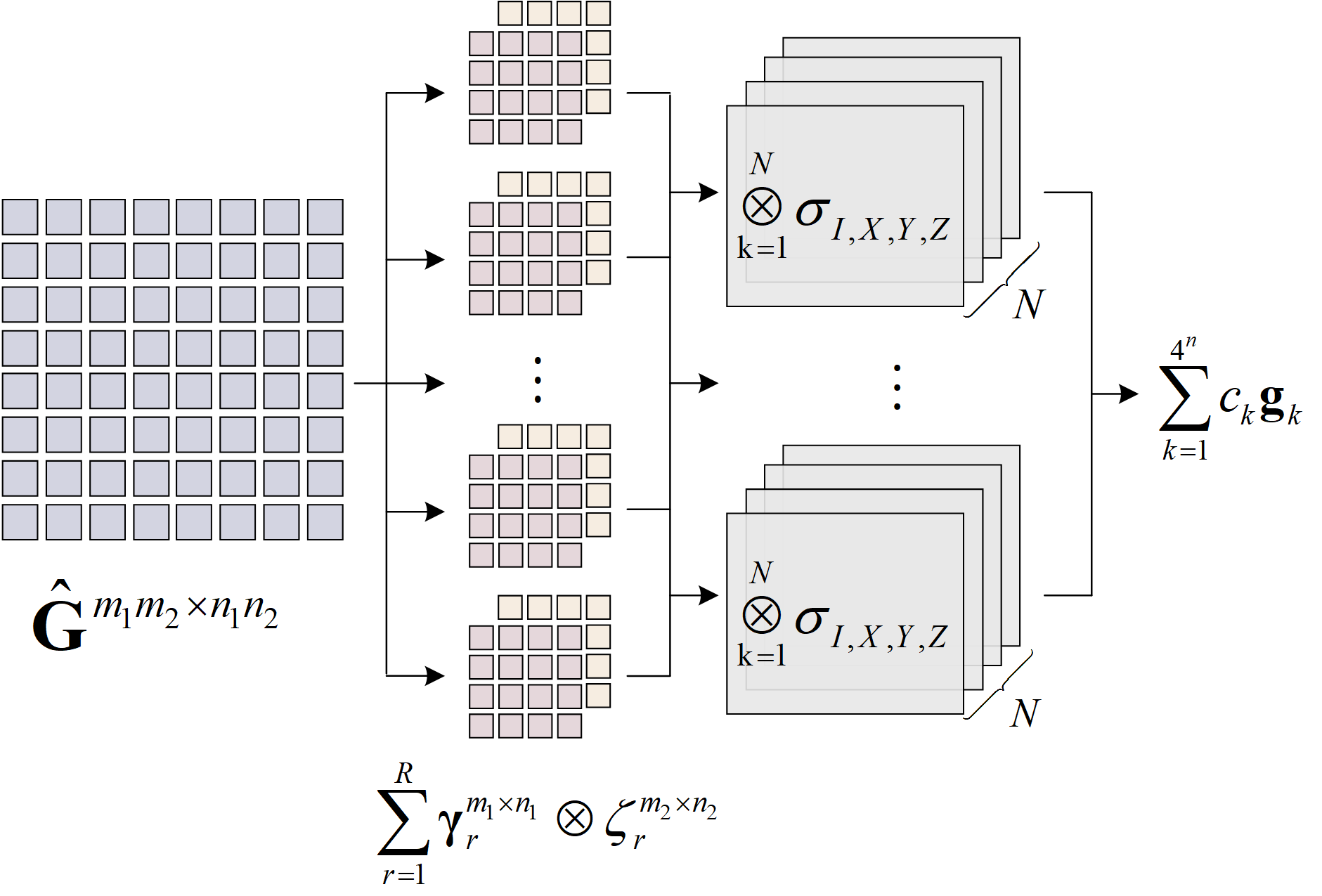}
    \setlength{\abovecaptionskip}{-2pt}
    \vspace{-0cm}
    \caption{The schematic diagram of the MLQC algorithm framework.
}
    \label{fig:MLQC}
  \end{center}
\end{figure}

Based on Theorem \ref{mapping1}, for any matrix, its mapping in the Pauli operator space is unique. Therefore, Since the ${\mathbf{\hat{G}}}$ obtained from the MLQC method is equal to ${\mathbf{\bar{\tilde{G}}}}$, the solution of MLQC is exactly the same as that in \eqref{ci_cal}. This is further illustrated in Fig. \ref{fig:MLQC}. 

\begin{remark}
    Here, it is worth mentioning that for an $n$-qubit circuit, a matrix of dimension $N=2^{n}$ requires $4^{n}$ multiplications of N-dimensional matrices through \eqref{ci_cal}. In our proposed MLQC, only $R\cdot2^{n+1}$ computations of $\sqrt{N}$-dimensional matrices are required. Currently, on classical computers, the computing complexity of matrix computation grows polynomially with the dimension of the matrix as $O(N^{p})$. Thus, the time complexity of the Pauli matrix mapping is $O(4^{\left \lceil  \log_{2}{N} \right \rceil}N^{p})$. In our proposed MLQC, thanks to the dimension reduction through GKD, exponential speedup is achieved. This will be verified in section \ref{case_study}.
\end{remark}

\subsection{Real-only Quantum Circuit Reduction Method}
In this section, we will present our proposed real-only quantum circuit reduction method that reduces the number of quantum circuits by half while maintaining the precision of the solution. 
\subsubsection{Y-Gate Even Principle}
Since the operator space constructed by Pauli matrices can represent any matrix in the complex space, while the admittance matrices in EMTP are often real and symmetric according to its physical characteristics \cite{TNE01Symmetric}, there is redundancy in the operator space constructed by Pauli operators. Therefore, we can prove Theorem \ref{oddY} for \textbf{the first time}, enabling the design of the real-only quantum circuit reduction method.

\begin{theorem}[Y-Gate Even Principle for Pauli Mapping of Real Symmetric Matrices]\label{oddY}
For $\forall \mathbf{\tilde{G}} \in  \mathbf{R}^{2^{n}\times 2^{n}}$, if $\mathbf{\tilde{G}}$ is a real symmetric matrix, the projection for the inner product of $\mathbf{\tilde{G}}$ on any basis in the Pauli space $\otimes_{k=1}^{n}\{\sigma_{I,X,Y,Z}\}$ that contains an odd number of $\sigma_{Y}$ must be zero.
\end{theorem}
\begin{proof}
    Let $\mathbf{\tilde{G}}$ be a real symmetric matrix with its projection mapping in the Pauli operator space as $\sum_{i=1}^{4^{n}}c_{i}\mathbf{g}_{i}, \mathbf{g}_{i}\in \otimes_{k=1}^{n}\{\sigma_{I,X,Y,Z}\}$, we have
    \begin{equation}\label{AAT}
    \mathbf{\tilde{G}}+\mathbf{\tilde{G}}^{T}=\sum_{i=1}^{4^{n}}c_{i}(\mathbf{g}_{i}+\mathbf{g}_{i}^{T}).
    \end{equation}
    
For each element of ($\mathbf{g}_{i}+\mathbf{g}_{i}^{T}$),  since $\sigma_{I,X,Z}$ are symmetric matrices, any $\mathbf{g}_{i}^{I,X,Z}$ which means $\mathbf{g}_{i}$ contains only $\sigma_{I,X,Z}$ will satisfy $\mathbf{g}_{i}^{I,X,Z}=(\mathbf{g}_{i}^{I,X,Z})^{T}$. For matrices involving $\sigma_{Y}$, we have
    \begin{equation}\label{YYT}
    \begin{split}
        \sigma_{Y}+\sigma_{Y}^{T}&=0\\(\sigma_{Y}\otimes\sigma_{I,X,Z}^{\otimes n}\otimes\sigma_{Y})^{T}&=\sigma_{Y}\otimes\sigma_{I,X,Z}^{\otimes n}\otimes\sigma_{Y}
    \end{split}
    \end{equation}
    Here, $\sigma_{I,X,Z}^{\otimes n}$ represents the combinations of tensor products involving $\sigma_{I,X,Z}$. It can be seen that any $\mathbf{g}_{i,even}$ containing an even number of $\sigma_{Y}$ is also symmetric, while for $\mathbf{g}_{i,odd}$ with an odd number of $\sigma_{Y}$, it yields
    \begin{equation}\label{Yodd}
        \mathbf{g}_{i,odd}=\sigma_{I,X,Y}^{\otimes n_{1}}\otimes\sigma_{Y}\otimes\sigma_{Y,enven}^{\otimes n_{2}}.
    \end{equation}
Here, $\sigma_{Y,enven}^{\otimes n_{2}}$ represents the tensor product combinations of Pauli matrices containing an even number of $\sigma_{Y}$ matrices. Then, we have
    \begin{equation}\label{YoddYT}
\mathbf{g}_{i,odd}+\mathbf{g}_{i,odd}^{T}=\sigma_{I,X,Y}^{\otimes n_{1}}\otimes(\sigma_{Y}+\sigma_{Y}^{T})\otimes\sigma_{Y,even}^{\otimes n_{2}}
        =0.
    \end{equation}
    
    By separating \eqref{AAT} based on the presence of $\sigma_{Y}$, we obtain
    \begin{equation}\label{AAT_re}
    \begin{split}
            \mathbf{\tilde{G}}+\mathbf{\tilde{G}}^{T}=\sum_{i}c_{i}(\mathbf{g}_{i}^{I,X,Z}&+(\mathbf{g}_{i}^{I,X,Z})^{T}\\
            +\sum_{j}(c_{j}\mathbf{g}_{j,even}&+(\mathbf{g}_{j,even})^{T})\\
             +\sum_{k}(c_{k}\mathbf{g}_{k,odd}&+(\mathbf{g}_{k,odd})^{T}).
    \end{split}
    \end{equation}
    
    Substituting \eqref{YoddYT} into \eqref{AAT_re} yields
    \begin{equation}\label{AATfinal}
        \begin{split}
             \mathbf{\tilde{G}}+\mathbf{\tilde{G}}^{T}&=\sum_{i}2c_{i}\mathbf{g}_{i}^{I,X,Z}+\sum_{j}2c_{j}\mathbf{g}_{j,even}+0\\
            &=\sum_{i=1}^{4^{n}}2c_{i}\mathbf{g}_{i}.
        \end{split}
    \end{equation}
    
    According to Theorem \ref{mapping1}, the matrix $(\mathbf{\tilde{G}}+\mathbf{\tilde{G}}^{T})$ has a unique projection. Therefore, the two expressions in \eqref{AATfinal} are equivalent. Consequently, for each element, $\mathbf{g}_{k,odd}$, its projection value must be zero.
\end{proof}


This theorem points to an important feature of the Pauli matrices that, among the four Pauli matrices, only $\sigma_{Y}$ is asymmetric. Therefore, any Pauli basis containing an odd number of $\sigma_{Y}$ matrices will be inherently asymmetric. 
\subsubsection{Quantum Circuit Reduction}
Then, referring to \eqref{had_local_cost}, all imaginary parts can be deleted. So, since $\sigma_{I,X,Z}$ contain no imaginary elements, each $\mathbf{g}_{i}$ is a real symmetric matrix, leading $\delta_{ii^{\prime}j}$ and $\beta_{ii^{\prime}}$ also to be real numbers. Therefore, we only need to calculate the real part of the Hadamard test result. Our method reduces half of the quantum circuits without loss of accuracy. 

More specifically, for the real symmetric matrix $\mathbf{\tilde{G}}$ in Theorem \ref{oddY}, the range for the number of effective basis elements, $\mathbb{N}_{c\neq 0}$, is given by\footnote{$\sum_{i=0}^{\lceil\frac{n}{2}\rceil}C_{n}^{2i}3^{n-2i}$ is the sum of the quantities of $\mathbf{g}_{i}^{I,X,Z}$ and $\mathbf{g}_{j,even}$ under \( n \) qubits.
}.
\begin{equation}\label{efficient basis calcu}
      2^{n}< \mathbb{N}_{c\neq 0} \le \sum_{i=0}^{\lceil\frac{n}{2}\rceil}C_{n}^{2i}3^{n-2i}
\end{equation}
Here, the lower bound $2^{n}$ refers to the case where the admittance matrix only contains diagonal elements, which, however, is not feasible since it implies that each node in the network is isolated. Therefore, in practice,  $\mathbb{N}_{c\neq 0}$ is closer to the upper bound. For MLQC-based QEMTP, the number of quantum circuits reduced in each cost function, denoted as $\mathscr{N}$, is given by
\begin{equation}\label{reduce_num}
\mathscr{N} = n\mathbb{N}_{c\neq 0}^{2}.
\end{equation}
For traditional VQLS, this number grows as $n4^{2n}$. In this way, the reduced number increases exponentially as the problem size grows. Note that this is just the amount that can be reduced in a single calculation. Considering that EMTP typically requires tens to hundreds of calculations, our proposed strategy can significantly reduce the number of quantum circuits. 
\begin{remark}
   For a problem of dimension \( N = 2^n \), traditional VQLS requires \( 2n4^{2n} \) quantum circuits to be computed in parallel. When \( n = 7 \), this requires more than $1$ billion quantum circuits needed for each cost function evaluation, making it impractical for quantum computers. In contrast, our method achieves substantial circuit reduction and holds even for a small number of qubits. The excellement performances can be referred to Table.\ref{table.rmse}.

\end{remark}

\subsubsection{Acceleration of the MLQC Procedure}
In addition, Theorem \ref{oddY} can also accelerate the MLQC computation. Van Loan proved in \cite{van1993approximation} that NKD can preserve the symmetry of matrices. Although there may be some asymmetry in GKD, in engineering practice, it is often less than half of the total number $R$. Therefore, for the symmetric part of the matrix in MLQC, it is only necessary to consider the effective basis up to the upper bound given in \eqref{efficient basis calcu}. This will further accelerate MLQC.

\subsection{The FASM for QEMTP}
Here, to further alleviate the computing burden caused by frequent changes in the admittance matrix as discussed in Section \ref{Matrix Projection}, we resort to an FASM by replacing switches with a fixed admittance equivalent circuit \cite{wang2018generalized}.

\begin{figure}[!htbp]
  \begin{center}
    \includegraphics[scale=0.45]{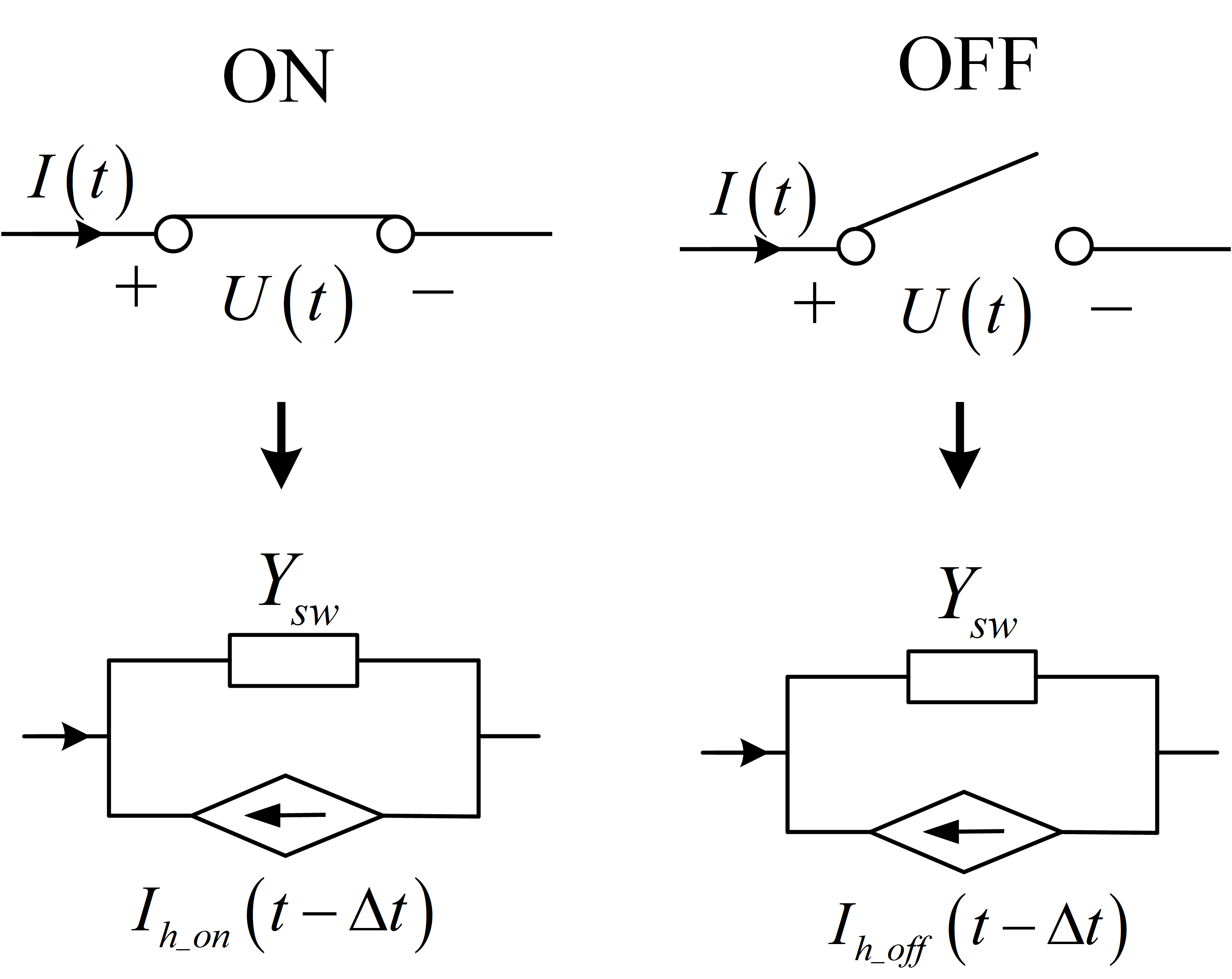}
    \setlength{\abovecaptionskip}{-0pt}
    \vspace{-0cm}
    \caption{The Framework of the FASM Equivalent Circuit.
}
    \label{fig:fasm}
  \end{center}
\end{figure}

This model was initiated in \cite{wang2018generalized} that utilizes a parameterized approach to represent the switch element directly as a Norton equivalent circuit in the form of an equivalent resistor in parallel with a current source, as shown in Figure \ref{fig:fasm}. Then, its current can be calculated as 

\begin{equation}\label{FASM}
    \begin{split}
        I_{h_{-}\text{on}}(t)&=\alpha_{\text {on }}Y_{\text{sw}} U(t-\Delta t)+\beta_{\text{on}} I(t-\Delta t) \\
        I_{h_{-}\text{off}}(t)&=\alpha_{\text {off }} Y_{\text{sw}}U(t-\Delta t)+\beta_{\text{off}} I(t-\Delta t).
    \end{split}
\end{equation}
Here, $\alpha_{\text {on }}$ and $\beta_{\text{on}}$ are the voltage and current coefficients at the switch conduction time, while $\alpha_{\text {off }}$ and $\beta_{\text{off}}$ are the voltage and current coefficients at the switch turn-off periods. Considering the equivalent model should exhibit the same steady-state characteristics as an ideal switch, the parameter relationships can be derived in 
\begin{equation}\label{FASM_para}
    \begin{split}
        \alpha_{\text {on }} &\neq 1, \beta_{\text {on }}=-1 \\
        \alpha_{\text {off }}&=1, \beta_{\text {off }} \neq-1.
    \end{split}
\end{equation}

For undetermined parameters $\alpha_{\text {on }}$ and $\beta_{\text {off }}$, different optimal damping parameters can be applied depending on the type of circuit to ensure that the FASM model has a rapidly decaying transient error. Its specific parameter settings for different circuits are provided in \cite{wang2018generalized}.
\begin{remark}
    Note that although FASM simplifies the EMTP model to some extent, we find that this is at least a practical strategy for QMETP that considers the power electronic switching actions. Using FASM, to the best of our knowledge, this is the first time that a QEMTP can be conducted with power electronic devices.  
\end{remark}

\subsection{Implementation Procedure}
In summary, Algorithm 1 presents the MLQC-based QEMTP procedure that we proposed.
\begin{algorithm}[!htbp]
\caption{The MLQC-based QEMTP Algorithm}
\begin{algorithmic} [1]
\State Choose the initial values for the parameters $T_{end}$, $\epsilon_{set}$ and system parameters;  
\State Constructing the admittance matrix $\mathbf{\tilde{G}}$ based on \textbf{FASM model};
\State $\triangleright \   $\textbf{\textit{MLQC procedure:}}
\State Decompose $\mathbf{\tilde{G}}$ by \eqref{G_decomp} and calculate the Pauli projection of each part by \eqref{G_decomp_cast};
\State Obtain the final projection result by \eqref{G_decomp_final};
\State $\triangleright \ $\textbf{\textit{Solve QEMTP equation \eqref{QEMTP_frame}}}:
\For{$t=0...,T_{end}-1$}
\State Compute $\vec{i(t)}$ by \eqref{EMTP} and prepare $|i\rangle$ by \eqref{vi-norm};
\State Construct the ansatz circuit as Fig.\ref{fig:ansatz} and initialize randomly;
\State Compute the cost function utilizing \textbf{Real-only Quantum Circuit Reduction Method};
\State Train to obtain the optimal parameters;

\State Measure and get $|v^{(0)}\rangle$ by quantum state tomography;
\State Compute $\epsilon^{(0)}$ by \eqref{compen_break} and $\Delta i^{(0)}$ by \eqref{delta_v};
\While{$\epsilon^{(k)}>\epsilon$}
\State Compute $\Delta v^{(k)}$ and update $|v^{(k)}\rangle$ by \eqref{compensate_alg};
\State Recompute $\epsilon^{(k)}$, $\Delta i^{(k)}$ and $k=k+1$;
\EndWhile
\EndFor
.

\end{algorithmic}\label{alg:HMS}
\end{algorithm}

\section{Case study}\label{case_study}
This section provides case studies on EMT networks containing multiple high-speed switching elements. The simulation results confirm the accuracy and losslessness of the proposed method.  Our experiments are based on the Origin Quantum Cloud Platform (PyQpanda) and are supported by the Big Data Computing Center of Southeast University.

\subsection{Verification of MLQC Speed Performance}
\subsubsection{Verification of the Speed and Accuracy of the MLQC Method}
Here, we analyze cases where the qubit number $n$ corresponds to the matrix dimensions $N=2^{n}$ that range from $2$ to $10$. By constructing random real symmetric matrices, we compare the projection computation time and accuracy of the MLQC method with that of the Pauli matrix mapping method. For lower qubit number ($2\sim 8$), we repeat each experiment $100$ times and take the average as the final result; for $n=9$ and $n=10$, due to the long computation time, we repeat the process $25$ times. Table \ref{table.MLQC} presents the computation times and mapping error $\epsilon$ for both methods along with the acceleration factor. Mapping error is quantified via
\begin{equation}\label{mapping_err}
    \varepsilon = \frac{\|\mathbf{\tilde{G}}-\sum_{i=1}^{4^{n}}2c_{i}\mathbf{\tilde{G}}_{i}\|_{F}}{\|\mathbf{\tilde{G}}\|_{F}}.
\end{equation}
\begin{figure}[!htbp]
  \begin{center}
    \includegraphics[scale=0.4]{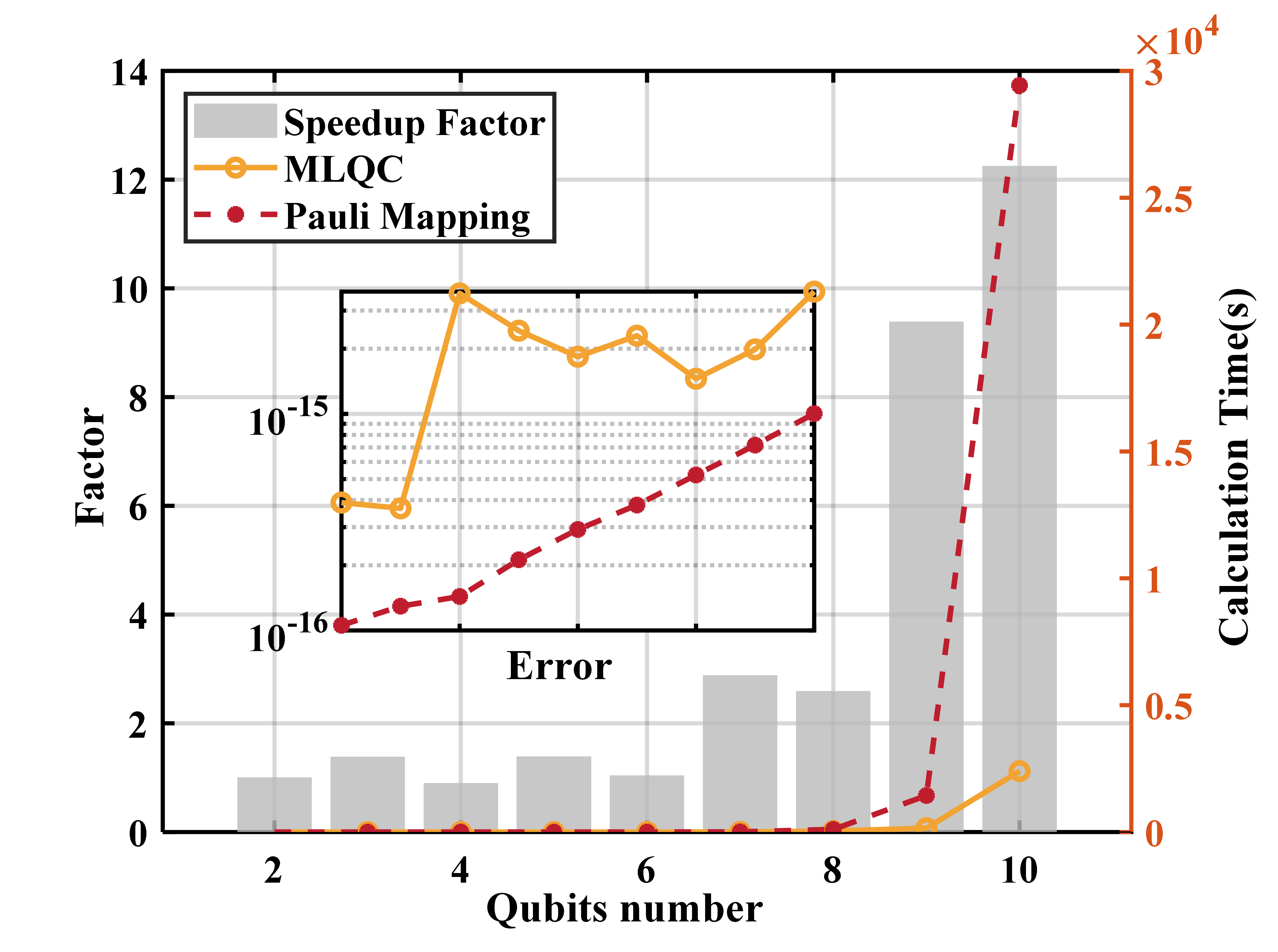}
    \setlength{\abovecaptionskip}{-0pt}
    \vspace{-0cm}
    \caption{Comparison of Computation Time Between MLQC and Pauli Mapping(R=$\min \{m_{1} n_{1}, m_{2} n_{2}\}$).
}
    \label{fig:MLQC_result}
  \end{center}
\end{figure}
Fig. \ref{fig:MLQC_result} presents a visual comparison of the time costs for the MLQC and Pauli mapping. The figure clearly shows that the acceleration effect of MLQC becomes more pronounced as the matrix size increases. For matrix dimensions of 512 (equivalent to 9 qubits) or larger, MLQC achieves a speedup of approximately tenfold or more. The figure illustrates that as the number of qubits increases, the acceleration factor increases exponentially. The detailed data is provided in Table \ref{table.MLQC} in Appendix \ref{MLQC_DATA}.
Furthermore, all mapping result errors are below $10^{-14}$, which is within the allowable range of the computing error. This shows that MLQC also preserves the accuracy of projection in a lossless manner while greatly accelerating the algorithm.
\subsubsection{The Impact of $R$ on the MLQC Method}
\begin{figure}[!htbp]
  \begin{center}
    \includegraphics[scale=0.4]{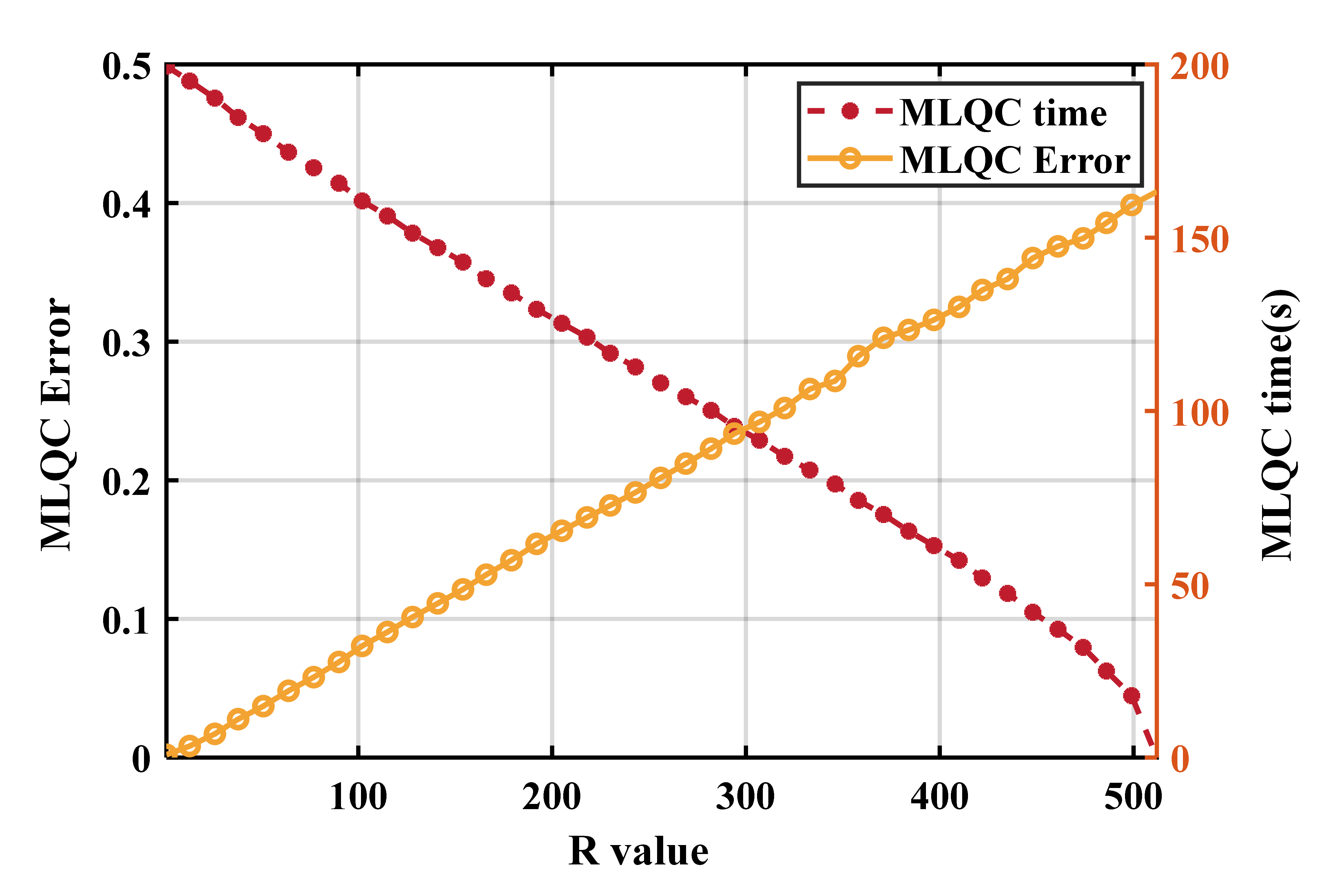}
    \setlength{\abovecaptionskip}{-0pt}
    \vspace{-0cm}
    \caption{MLQC results for different R values when qubits number is 9.
}
    \label{fig:MLQC_R_change}
  \end{center}
\end{figure}

The results in Fig. \ref{fig:MLQC_result} are obtained when R is set to the maximum value $\min \{m_{1} n_{1}, m_{2} n_{2}\}$. Next, we will discuss the case where R takes smaller values. As $R$ decreases, the number of low-dimensional matrix pairs mapped by MLQC decreases, and the subsequent computation speed of the projection increases. However, the trade-off is a reduction in computational accuracy. Fig. \ref{fig:MLQC_R_change} shows the variation of the MLQC results with respect to R when the number of qubits \(n=9\).

It can be seen that when \( R = 1 \), the MLQC has an accuracy error close to $0.5$. This means that each element of the matrix has an approximate deviation $50\%$, which is intolerable in practice. As \( R \) increases, both the overall computation time and the error decrease linearly. Therefore, in practical applications of MLQC, the maximum value of \( R \) is suggested. The remaining sections of this chapter also use \( R = \min \{m_{1} n_{1}, m_{2} n_{2}\} \) for the calculations.

\subsubsection{Special Property of the MLQC Method}
In \eqref{ge_Kronecker_eq}, considering the case where $R=1$, we can obtain the optimal rank-1 approximation of matrix $\mathbf{\tilde{G}}$. Therefore, in \( \mathbf{\tilde{G}}v = i \), if \( \mathbf{\tilde{G}} = \mathbf{\tilde{G}}_1 \otimes \mathbf{\tilde{G}}_2 \) and \( i = i_1 \otimes i_2 \), based on the Kronecker product properties, we have
\begin{equation}\label{Decomp_solve}
    \mathbf{\tilde{G}_1}v_1\otimes \mathbf{\tilde{G}_2}v_2  = i_1\otimes i_2.
\end{equation}

Based on \eqref{Decomp_solve}, the original problem \( \mathbf{\tilde{G}}v = i \) can be decomposed into two subequations \( \mathbf{\tilde{G}}_1 v_1 = i_1 \) and \( \mathbf{\tilde{G}}_2 v_2 = i_2 \). By solving these two equations separately, the solutions hold for \( v = v_1 \otimes v_2 \). This reduces the dimension in solving the original problem. However, this solution, which only uses NKD\footnote{When $R=1$, GKD is equivalent to NKD.}, often has low decomposition accuracy. In contrast, for the lossless GKD, we have \eqref{GKD_solve}.
\begin{equation}\label{GKD_solve}
\mathbf{\tilde{G}}v=\sum_{ij}\mathbf{\tilde{G}}_{1i}v_{1j}\otimes\mathbf{\tilde{G}}_{2i}v_{2j}
\end{equation}

Similarly to the case where \( R = 1 \), decomposing \( i \) as \( \sum_k i_{1k} \otimes i_{2k} \) suggests that a similar decomposition of subequations could be obtained. However, since the number of created subequations is much larger than the number of variables in the decomposition of \( v \), there is a coupling between the subequations, preventing them from being solved independently. This issue hinders further exploitation of the MLQC method.

\subsection{Simulation Results of High-Speed Switch DC-DC EMT Network}
\begin{figure*}[!htbp]
  \begin{center}
    \includegraphics[scale=0.3]{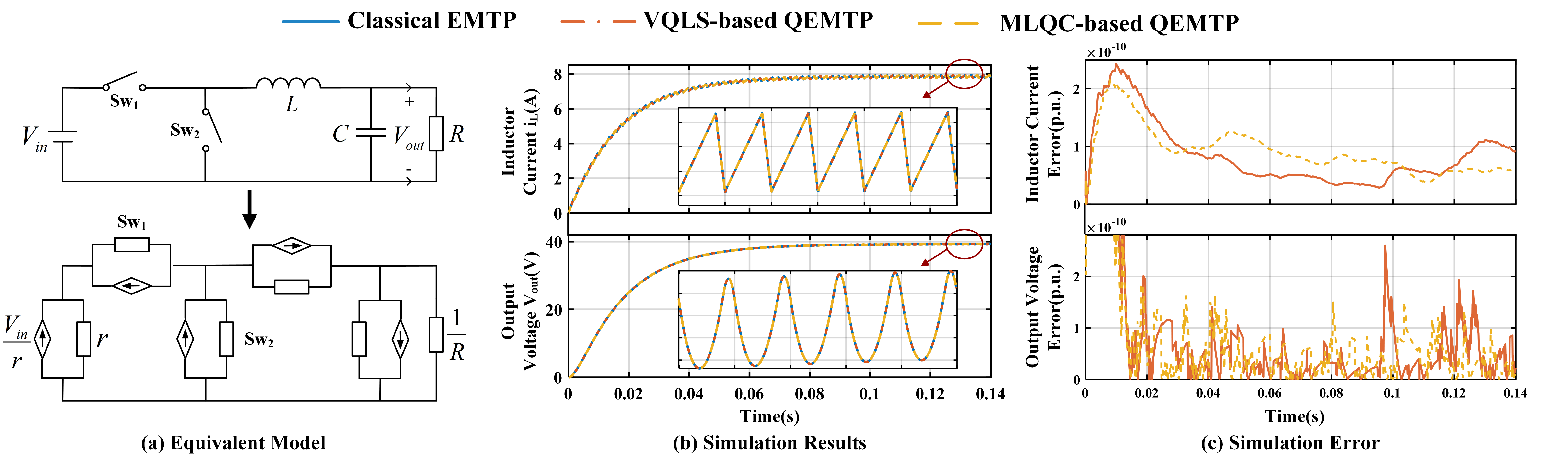}
    \setlength{\abovecaptionskip}{-0pt}
    \vspace{-0cm}
    \caption{Basic topology of the Buck circuit and simulation results.
}
    \label{fig:Buck_sim}
  \end{center}
\end{figure*}

To validate the simulation accuracy of the MLQC-based QEMTP method in DC-DC EMT networks, a Buck circuit containing a pair of alternating switches was selected.

\subsubsection{Parameter Settings}
The network structure of the Buck circuit is shown in Fig. \ref{fig:Buck_sim}. (a), with an input voltage of $50V$ and a duty cycle of 0.8. For FASM, the optimal parameters given in \cite{wang2018generalized} were chosen to minimize the transient process, as shown in Table \ref{table.para}. 
\begin{table}[!htbp]
\renewcommand{\arraystretch}{1.5}
\caption{ Best Damping Parameters (Shortest Transient Response)}
\label{table.para}
\centering
\begin{tabular}{ccccc}
\hline\hline
  \makecell[c]{$\alpha_{\text {on }}$} & \makecell[c]{$\beta_{\text {on}}$}
  &\makecell[c]{$\alpha_{\text {on }}$}&\makecell[c]{$\alpha_{\text {on }}$}
  &\makecell[c]{$Y_{\text{sw}}$}
  \\
\hline
$-1-\sqrt{2}$  & $-1$ & $1$ & $1-\sqrt{2}$ &$\sqrt{C/L}$ \\
\hline\hline
\end{tabular}
\end{table}

The QEMTP accuracy threshold $\epsilon$ is set to $10^{-7}$, and the number of ansatz layers is set to 3. Considering that the threshold defined in \eqref{compen_break} represents an absolute error limit that evaluates the overall computational precision, it is less effective to evaluate the precision of specific elements. Therefore, for a specific variable $v$ with a true value $v^{*}$, a relative error limit is proposed as 

\begin{equation}\label{err_r}
    \epsilon_{r}=\|\frac{v-v^{*}}{v^{*}}\|.
\end{equation}
The final simulation results are presented in Fig. \ref{fig:Buck_sim}(b), which show the operation of the Buck circuit under continuous current mode. Fig. \ref{fig:Buck_sim}(c) presents the error analysis for each time step. 

\subsubsection{Accuracy Verification of MLQC-based QEMTP}
\begin{figure*}[!htbp]
  \begin{center}
    \includegraphics[scale=0.3]{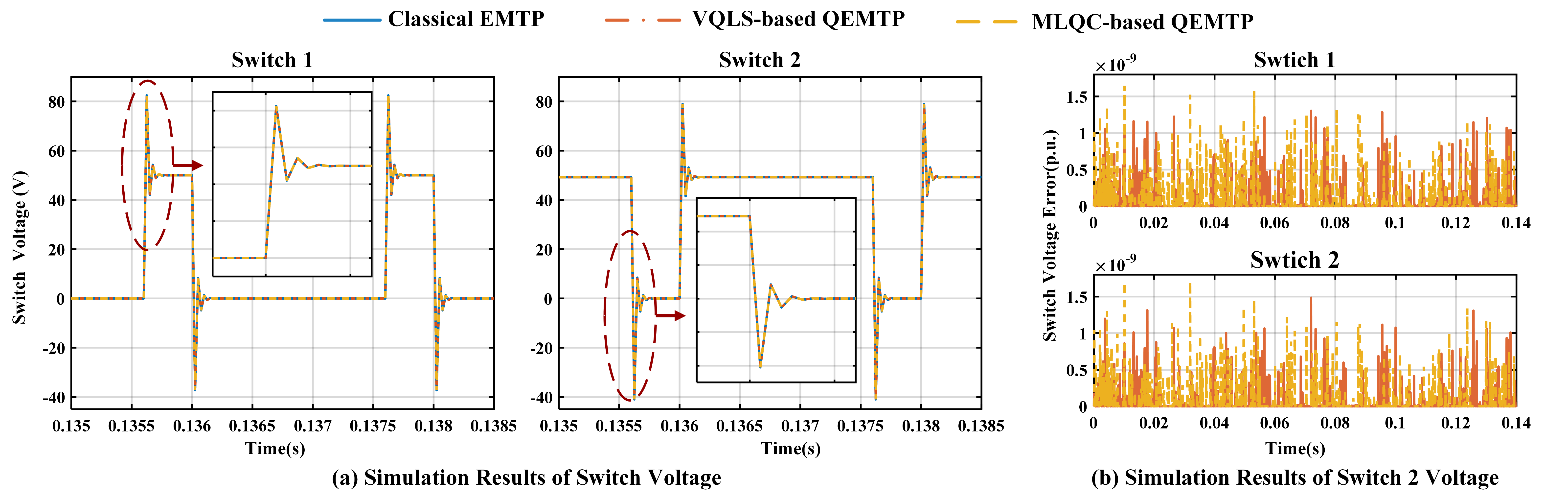}
    \setlength{\abovecaptionskip}{-0pt}
    \vspace{-0cm}
    \caption{Simulation results of switch voltages in Buck circuit.
}
    \label{fig:Buck_switch}
  \end{center}
\end{figure*}
 For FASM, we present the voltage waveforms of switches 1 and 2 in Fig. \ref{fig:Buck_switch}. 
  Although there is some transient fluctuation due to the initial conditions of the model, the system converges quickly within the $4-5$ time steps. At the switching points, the errors of both quantum methods and the classical EMTP remain within $10^{-9}$, which meets the set accuracy threshold. Most relative errors of both methods are around $10^{-9}$ and $10^{-10}$, and there is no error accumulation. This shows that the MLQC-based QEMTP method can achieve high-precision simulations for DC-DC EMT networks with high-speed switching elements. Moreover, compared to the VQLS-based QEMTP, there is no significant difference in error accumulation.

\subsubsection{Quantum Circuit Reduction}
In this case, MLQC-based QEMTP saves 128 quantum circuits in each cost function calculation. As a 4-dimmension case, each time step requires approximately 70 iterations, with 3 iterations for corrections. As a result, 26,880 quantum circuits will be saved in each time step. Therefore, in the 0.14s simulation process with a time step of $25\mu s$, our method can reduce about 150.53 million quantum circuits.

\subsection{Simulation Results of High-Speed Switch AC-DC EMT Network}
\begin{figure}[!htbp]
  \begin{center}
    \includegraphics[scale=0.4]{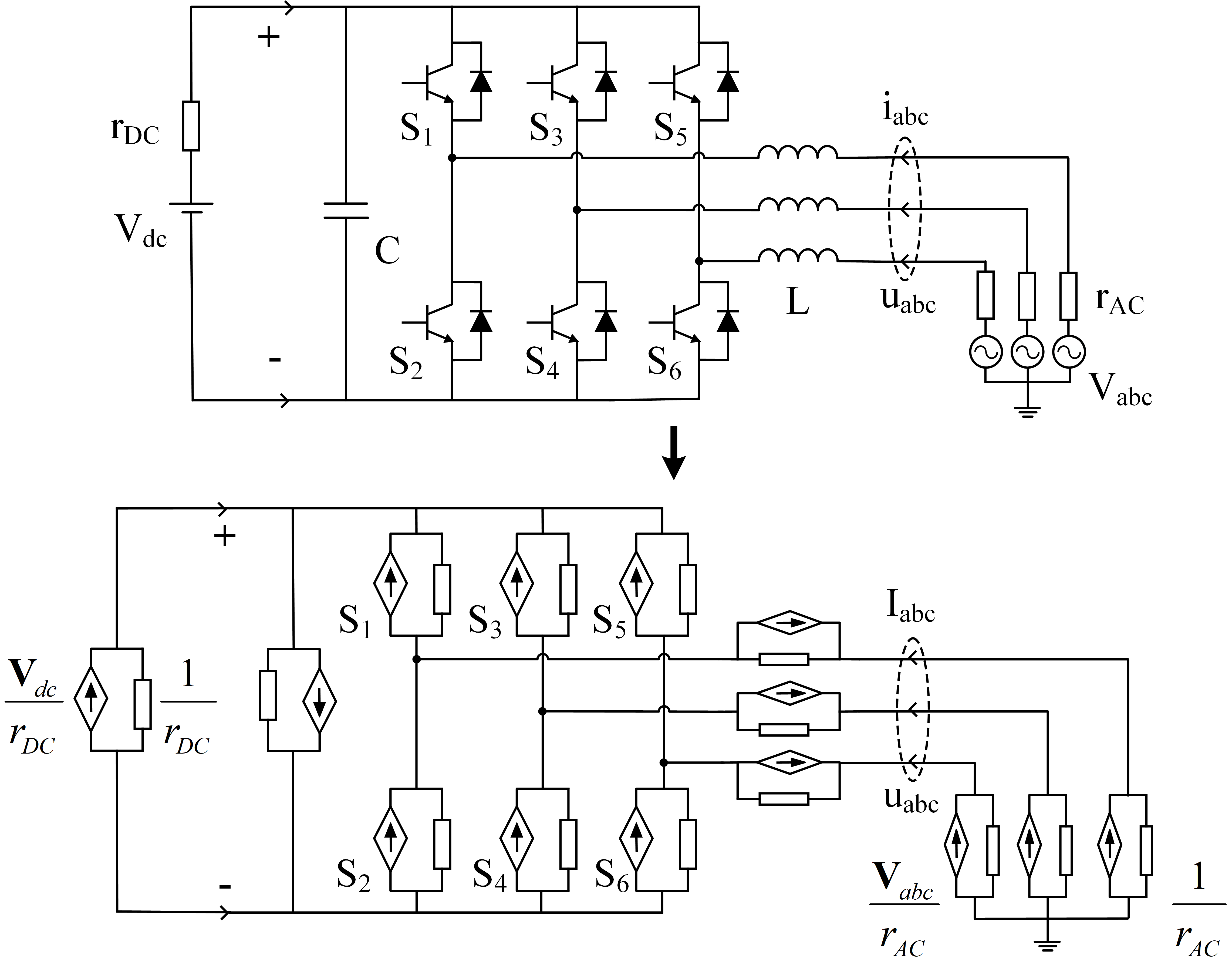}
    \setlength{\abovecaptionskip}{-0pt}
    \vspace{-0cm}
    \caption{The framework of three-phase full bridge inverter.
}
    \label{fig:convertor}
  \end{center}
\end{figure}

To validate the accuracy of the MLQC-based QEMTP in AC EMT networks, a three-phase full-bridge rectifier, as shown in Fig.\ref{fig:convertor}, is selected as the simulation case. 
\subsubsection{Parameter Settings}
Here, the AC grid side is connected to three pairs of alternately conducting switches, controlled by PWM signals to achieve AC-DC rectification. The triangular carrier frequency is set to 2.5 kHz, the resistance to side load DC is $50\Omega$, and the phase voltage of the AC grid is $326.6 V$. The simulation time step is 10 $\mu$s.

Considering the higher AC-side voltage and the more complex circuit structure, the QEMTP accuracy threshold $\epsilon$ is set to $10^{-4}$, and the number of ansatz layers is set to 5.
\begin{figure}[!htbp]
  \begin{center}
    \includegraphics[scale=0.5]{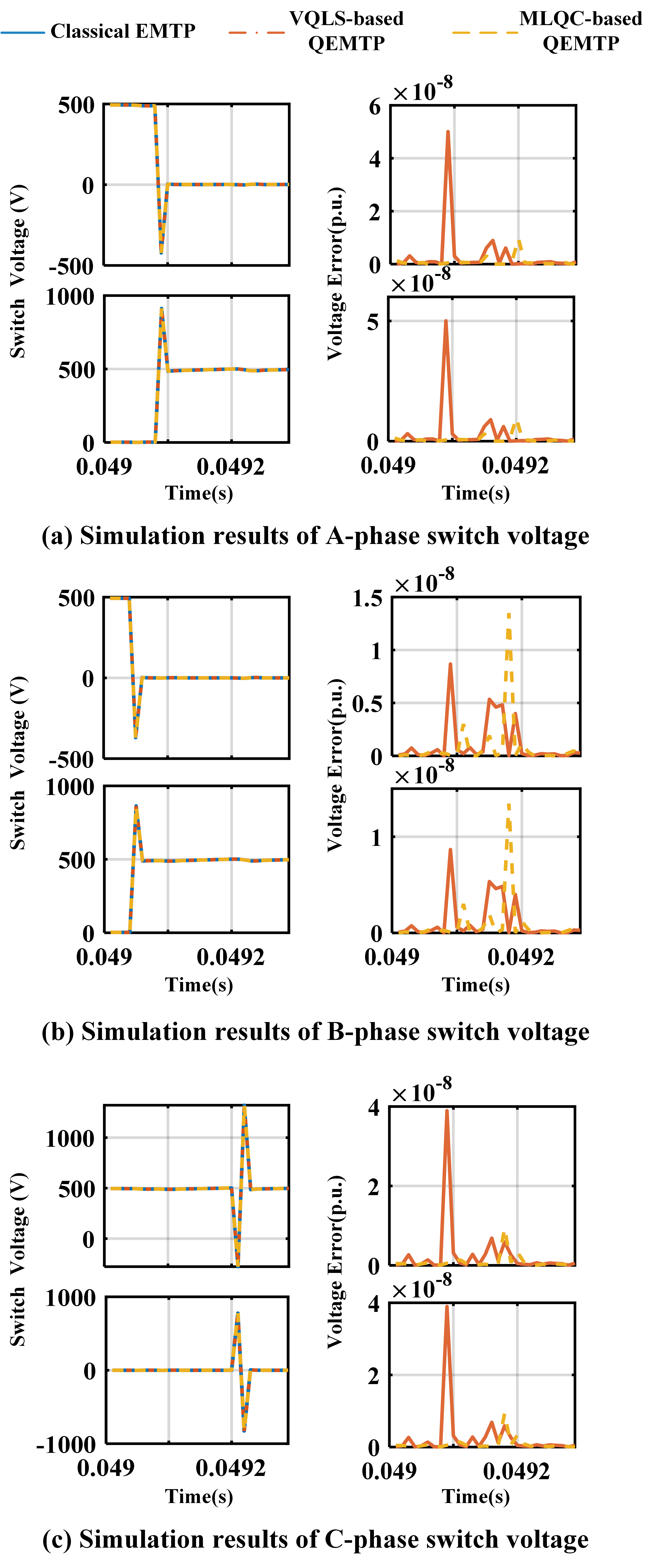}
    \setlength{\abovecaptionskip}{-0pt}
    \vspace{-0cm}
    \caption{Simulation results of switch voltages in three-phase full bridge inverter.}
    \label{fig:convertor_result}
  \end{center}
\end{figure}

\subsubsection{Verification of MLQC-based QEMTP}
Fig. \ref{fig:convertor_result} shows the simulation waveforms of switching voltage of a three-phase full-bridge inverter. Due to resource limitations on the original quantum cloud platform, only a simulation from 0.049 to 0.0493 seconds was computed. The simulation results indicate that the error is within $10^{-7}$, with no observable error accumulation. Furthermore, there are no significant differences in errors between the two QEMTP methods. This shows that our method can achieve accurate simulation results even in more complex AC-DC networks.
\subsubsection{Quantum Circuit Reduction}
In this 8-dimmension case, each cost function calculation in the MLQC-based QEMTP saves 2,700 quantum circuits. Meanwhile, each time step requires approximately 300 iterations and 4 correction steps. As a result, for each time step, our method saves 3.24 million quantum circuits. For a 1 ms simulation with a time step of $10 \mu s$, more than 300 million quantum circuits can be saved.
Table.\ref{table.rmse} presents a comparison of the computational results for two cases. It can be observed that, while maintaining simulation accuracy, the number of quantum circuits saved by the Real-only construction criterion increases exponentially with the problem size. Even for small-scale examples, significant savings in quantum resources are achieved.
\begin{table}[!htbp]
\renewcommand{\arraystretch}{1.5}
\caption{ Comparison between two cases}
\label{table.rmse}
\centering
\begin{tabular}{c|cc|cc}
\hline
\hline
      & \multicolumn{2}{c|}{RMSE($10^{-9}$p.u.)} & \multicolumn{2}{c}{MLQC Quantum Circuit Reduction}                                       \\ \cline{2-5} 
      & \multicolumn{1}{c|}{VQLS}     & MLQC     & \multicolumn{1}{l|}{per step($10^{3}$)} & \multicolumn{1}{l}{1ms simulation($10^{6}$)} \\ \hline
DC-DC & \multicolumn{1}{c|}{0.1177}   & 0.09987  & \multicolumn{1}{c|}{26.88}              & 1.072                                          \\ \hline
AC-DC & \multicolumn{1}{c|}{8.9428}   & 4.8341   & \multicolumn{1}{c|}{3,240}              & 324                                            \\ \hline\hline
\end{tabular}
\end{table}

\section{Conclusions}\label{conclu}
This paper proposes an MLQC-based QEMTP method and validates it for the first time in power electronic networks featured multiple high-speed switching devices. 
The simulation results reveal excellent performance of the proposed method. 
This approach not only advances the practical application of QEMTP but serves as a versatile method with potential applications for simulating and solving various types of networks.


\begin{appendices}
\renewcommand{\theequation}{A.\arabic{equation}}
\setcounter{equation}{0}

\section{NKD Solution Process}\label{append:NKD}
In Section.\ref{Sec:NKD}, we presented the objective function \eqref{Kronecker_app} for NKD. In this section, we will provide a detailed solution. By partitioning the matrix into smaller submatrices, $\mathbf{\tilde{G}}$ can be decomposed as
\begin{equation}\label{matrix_eq}
\mathbf{\tilde{G}}=\left(\begin{array}{cccc}
\mathbf{\tilde{G}}_{11} & \mathbf{\tilde{G}}_{12} & \cdots & \mathbf{\tilde{G}}_{1, n_{1}} \\
\mathbf{\tilde{G}}_{21} & \mathbf{\tilde{G}}_{22} & \cdots & \mathbf{\tilde{G}}_{2, n_{1}} \\
\vdots & \vdots & \ddots & \vdots \\
\mathbf{\tilde{G}}_{m_{1}, 1} & \mathbf{\tilde{G}}_{m_{1}, 2} & \cdots & \mathbf{\tilde{G}}_{m_{1}, n_{1}}
\end{array}\right).
\end{equation}
Here, $\mathbf{\tilde{G}}_{ij}\in \mathbf{R}^{m_{2}\times n_{2}}$. An equivalent form of \eqref{Kronecker_app} can be rearranged as \cite{van1993approximation}
\begin{equation}\label{Kronecker_eq}
\|\mathbf{\tilde{G}}-\mathbf{B} \otimes \mathbf{C}\|_{F}=\left\|\bar{\mathbf{\tilde{G}}}-\operatorname{vec}(\mathbf{B}) \operatorname{vec}(\mathbf{C})^{T}\right\|_{F}.
\end{equation}
Here, $\operatorname{vec}(\mathbf{B})$ is the vectorized matrix $\mathbf{B}$ by stacking its columns, $\operatorname{vec}(\mathbf{C})$ is the same. ${\mathbf{\bar{\tilde{G}}}}$ is the permuted transformation of $\mathbf{\tilde{G}}$ as

\begin{equation}\label{A_tilde}
\mathbf{\bar{\tilde{G}}}=\left[\begin{array}{c}
\operatorname{vec}\left(\mathbf{\tilde{G}}_{11}\right)^{T} \\
\operatorname{vec}\left(\mathbf{\tilde{G}}_{21}\right)^{T} \\
\dots  \\
\operatorname{vec}\left(\mathbf{\tilde{G}}_{m_{1}n_{1}}\right)^{T}.
\end{array}\right]
\end{equation}

Assume that the singular value decomposition (SVD) of the matrix ${\mathbf{\bar{\tilde{G}}}}$ is given by $\bar{\mathbf{\tilde{G}}}=\sum_{r=1}^{\min \{m_{1} n_{1}, m_{2} n_{2}\}} \sigma_{r} \boldsymbol{u}_{r} \boldsymbol{v}_{r}^{T}$
 , where $\sigma_{1}>\sigma_{2}>\dots>\sigma_{\min \{m_{1} n_{1}, m_{2} n_{2}\}}$, then the optimal solutions for matrices $\mathbf{B}$ and $\mathbf{C}$ are provided by 

\begin{equation}\label{naive_ans}
\begin{split}
    \operatorname{vec}(\mathbf{B})&=\sqrt{\sigma_{1}}\boldsymbol{u}_{1}\\
    \operatorname{vec}(\mathbf{C})&=\sqrt{\sigma_{1}}\boldsymbol{v}_{1}.
\end{split}
\end{equation}
\section{RESULT OF PAULI MAPPING AND MLQC}\label{MLQC_DATA}
In Fig.\ref{fig:MLQC_result}, we present a comparison of the mapping times between MLQC and Pauli-mapping. The specific data corresponding to the figure is provided below.
\begin{table}[!htbp]
\renewcommand{\arraystretch}{1.3}
\caption{ Result of Pauli Mapping and MLQC}
\label{table.MLQC}
\centering
\begin{tabular}{cccccc}
\toprule
\multirow{2}{*}{\begin{tabular}[c]{@{}c@{}}Matrix\\ Size\end{tabular}} & \multicolumn{2}{c}{MLQC} & \multicolumn{2}{c}{Pauli Mapping} & \multirow{2}{*}{\begin{tabular}[c]{@{}c@{}}Speedup\\ Factor\end{tabular}} \\ \cline{2-5}
 & time(s)& error(E-15)& time(s) & error(E-16)  &  \\ \midrule
4    & 0.00055 & 0.38961 & 0.000556 & 1.05638 & 1.005 \\
8    & 0.00242 & 0.36640 & 0.003342 & 1.29437 & 1.383 \\
16   & 0.02031 & 3.60241 & 0.01835  & 1.43524 & 0.903 \\
32   & 0.07413 & 2.42259 & 0.103213 & 2.12148 & 1.392 \\
64   & 0.75287 & 1.83753 & 0.784744 & 2.92788 & 1.042 \\
128  & 2.80550 & 2.29698 & 8.100441 & 3.79544 & 2.887 \\
256  & 39.9338 & 1.45355 & 103.6693 & 5.22249 & 2.596 \\
512  & 153.327 & 1.98560 & 1439.483 & 7.17661 & 9.388 \\
1024 & 2402.45 & 3.67004 & 29431.47 & 10.0334 & 12.25\\
\bottomrule
\end{tabular}
\end{table}

\end{appendices}
 \vspace{-0.3cm}
 \newcommand{\BIBdecl}{\setlength{\itemsep}{0.01 em}}
 \bibliographystyle{IEEEtran}
\bibliography{IEEEabrv,Reference.bib}

\vspace{-0.3cm}
\end{document}